\documentclass{emulateapj}
\usepackage{graphicx}
\usepackage{amssymb}
\usepackage{natbib}
\bibpunct{(}{)}{;}{a}{}{,}

\shorttitle{Evolution and proper motions of radio Source I and BN}
\shortauthors{Goddi et al. 2010}

\newcommand{\kms}{km~s$^{-1}\,$}
\newcommand{\vlsr}{$V_{\rm LSR}$}

\newcommand{\ms}{M$_{\odot}$}
\newcommand{\ls}{L$_{\odot}$}
\newcommand{\msyr}{M$_{\odot}$~yr$^{-1}$}
\newcommand{\pas}{$\rlap{.}^{\prime\prime}$}

\begin{document}

\title{A Multi-Epoch Study of the Radio Continuum Emission of
    Orion Source~I: Constraints on the Disk Evolution of a Massive YSO
  and the Dynamical History of Orion BN/KL
}

\author{C. Goddi, E. M. L. Humphreys}
\affil{European Southern Observatory, Karl-Schwarzschild-Strasse 2,
D-85748 Garching bei M$\ddot{u}$nchen}
\affil{Harvard-Smithsonian Center for Astrophysics,
    60 Garden Street, Cambridge, MA 02138}

\author{L. J. Greenhill}
\affil{Harvard-Smithsonian Center for Astrophysics,
    60 Garden Street, Cambridge, MA 02138}

\and

\author{C. J. Chandler}
\affil{National Radio Astronomy Observatory, P.O. Box O, Socorro, NM 87801}

\and

\author{L. D. Matthews}
\affil{MIT Haystack Observatory, Westford, MA 01886}

\begin{abstract}
We present new $\lambda7$~mm continuum observations of Orion BN/KL with the Very Large Array (VLA). 
We resolve the emission from the Young Stellar Objects (YSO) radio Source~I and BN at several epochs.
Radio Source~I is highly elongated northwest-southeast, and remarkably stable in flux density, position angle, and overall morphology over nearly a decade. This favors the extended emission component arising  from an ionized edge-on disk rather than an outwardly propagating jet.

We have measured the proper motions of Source~I and BN for the first time at 43~GHz. We confirm that both sources are moving at high speed (12 and 26~\kms, respectively) approximately in opposite directions, as previously inferred from measurements at lower frequencies.  
We discuss dynamical scenarios that can explain the large motions of both BN and Source~I and the presence of disks around both. Our new measurements support the hypothesis that a close ($\sim 50$~AU) dynamical interaction occurred around 500~years ago  between Source~I and BN as proposed by Gomez et al.  
%Our simulations show that a multi-body interaction among unbound objects has low probability to eject massive stars and form massive binaries. 
From the dynamics of encounter we argue that Source I today is likely to be a binary with a total
mass on the order of 20~\ms, and that it probably existed as a
softer binary before the close encounter.  This enables preservation
of the original accretion disk, though truncated to its present radius
of  $\sim 50$~AU. 

N-body numerical simulations show that the dynamical interaction between a binary of 20~\ms~total mass (Source~I) and a single star of 10~\ms~mass (BN) may lead to the ejection of both and binary hardening. The gravitational energy released in the process would be large enough to power the wide-angle, high-velocity flow traced by H$_2$ and CO emission in the BN/KL nebula. 
Assuming the proposed dynamical history is correct, the smaller mass for Source~I  recently estimated from SiO maser dynamics ($\gtrsim7$~\ms) by Matthews et al., suggests that  non-gravitational forces (e.g. magnetic) must play an important role in the circumstellar gas dynamics.
 
\end{abstract}


%===========================
\section{Introduction}
%===========================
\label{int}
The Orion~BN/KL complex, at a distance of $418\pm6$~pc \citep{Men07,Kim08}, contains the nearest region of ongoing high-mass star formation. 
A dense protostellar cluster lies within the region containing three
radio sources that are believed to be massive young stellar objects
(YSOs):  the highly embedded radio Source~I,
\citep{Gre04a,Rei07,Mat10}; 
the BN object, which is the brightest source in the region in the
mid-infrared (IR) 
at 12.4~$\mu$m \citep{Gez98}; and Source {\it n}, a relatively evolved
YSO with a disk observed in the MIR \citep{Gre04b} and a jet observed
in the radio at 8.4~GHz \citep{Men95,Gom08}. 

Despite intensive investigations at radio and IR wavelengths, the
primary heating source(s) for the Orion~KL region ($L\sim 10^5$~\ls)
is (are) still not known. 
Another long-standing puzzle is the geometry of  outflow and the
identification 
of driving sources. There are two large-scale outflows in the region. 
A powerful ($3\times 10^{47}$~ergs), high-velocity (30$-$200~\kms),
wide-angle ($\sim 1$~rad) outflow extends northwest-southeast (NW-SE) over $0.3$~pc.  This so-called ``high-velocity'' outflow is traced  in CO emission \citep{Che96,Zap09} and in 2.12~$\mu$m H$_2$ shocked emission originating in finger-like structures that end in bow shocks \citep{All93}. % The by product of an explosion through fragmentation of colliding flows has also been proposed.
A second, ``low-velocity'' outflow ($\sim 18$~\kms) is identified by a
cluster of bright $v=0$~SiO and H$_2$O masers, concentrated within a
few arcsec around Source~I and elongated northeast-southwest (NE-SW;
\citealt{Gen89,Gre98}, and in prep.).
 
Source~I has been proposed as a possible driver of both the
high-velocity NW-SE outflow (e.g., Wright et al. 1995;
\citealt{Gre98,Bal05}) and the low-velocity NE-SW outflow
(\citealt{Beu05,Gre04a} and in prep.). Confusion arises because the radio continuum emission from Source~I shows an elongated structure, which has been interpreted as both an ionized jet along a NW-SE direction \citep{Bal05,Tan08a} and as an ionized disk with a NE-SW spin axis \citep{Rei07}. 

%IN BRIEF: High-angular resolution observation observations have shown that the 7~mm continuum emission is highly elongated and is consistent with an edge-on disk, lying at the center of an X-shaped distribution of SiO masers in a disk and outflow (Matthews et al., Greenhill et al. 2009).

Based on a multi-epoch observational campaign of several SiO maser
transitions using the Very Large Array (VLA) and the Very Long
Baseline Array (VLBA), \citet{Mat10} and Greenhill et al. (in prep.) 
provide convincing evidence that Source~I is associated with a disk/outflow system with a NE-SW axis. In particular, based on a VLBA monitoring of  $v=1,2 \ J=1-0$ SiO maser transitions, \citet{Mat10} presented a movie of bulk gas flow tracing a compact disk and the base of a protostellar outflow at radii $<100$~AU from Source I. In addition, Greenhill et al. measured proper motions of $v=0$~SiO masers, which trace the bipolar outflow expanding with a characteristic velocity $\sim 18$~km~s$^{-1}$ and extending to radii of 100-1000\,AU from Source~I along a NE-SW axis, parallel to the axis of the disk/outflow system  traced by $v=1,2$ masers at radii $<100$~AU. 

The origin and nature of the wide-angle NW-SE oriented outflow, traced in shocked H$_2$ and CO emission, is still a matter of debate. \citet{Rod05} proposed that  Source~I and BN had a close encounter about 500~yrs ago, which resulted in the ejection of interacting sources and the formation of a tight binary (Source~I).  Based on a proper motion study of radio sources at 8.4~GHz, \citet{Gom08} proposed that Source~I, BN, and source {\it n} participated in the same dynamical interaction 500~yrs ago and that all three sources are moving away from the putative center of interaction. 
The energy liberated in the process may provide in principle sufficient energy to power the fast NW-SE outflow \citep{Bal05,Zap09}.
 It is not clear, however, what effect a close passage and consequent formation of a tight binary would have on a well-organized accretion/outflow structure such as observed in Source~I \citep{Mat10}. 
 \citet{Tan04,Tan08b} proposed an alternatively scenario where a close passage between Source~I and BN (a runaway star from the Trapezium) would trigger a tidally-enhanced accretion and subsequent outburst of outflow activity, resulting in the powerful high-velocity outflow.
 
 In this paper, we present new multi-epoch, high-angular resolution observations of the radio continuum emission in Orion BN/KL at 7~mm (43~GHz) from the VLA. The main goals of the new observations were to investigate  the nature of the radio continuum in Source~I and reconstruct the dynamical history of BN/KL.
%Our analysis is based on sensitive new multi-epoch, high-angular resolution observations of the radio continuum emission from Orion BN/KL at 7~mm (43~GHz) from the VLA. %We note that, as compared with previous 8.4~GHz continuum studies, observations at higher frequencies at 43~GHz, present a two-fold advantage, in that they enable higher angular resolutions together with the possibility of using the strong SiO maser signal as a phase-reference for calibration purposes \citep{Rei07}.
In particular, we investigate changes in morphology, size, and flux
density as a function of time (over a decade) and frequency (43 vs
8.4~GHz) to obtain insights into the nature of the radio sources,
mainly to test the  ionized disk and jet hypotheses proposed for
Source~I. In addition, we measured absolute proper motions of radio
sources based on accurate absolute astrometry, with the aim of
constraining the dynamical history of the BN/KL region. In order to
quantify probabilities of different dynamical scenarios, we present
also new N-body simulations of decaying protostellar clusters with a
varying number of objects. Previous N-body simulations for BN/KL
\citep{Gom08} assumed a five-member cluster having large masses (in
the range 8-20~\ms), which resulted in the formation of binaries with
total mass of 36~\ms. However, there is no evidence of such massive
objects in the BN/KL region, based on present data. Our new
simulations assume more plausible masses of individual objects as well
as investigate a larger number of possible scenarios.

The current paper is structured as follows.
The  observational setup and data calibration procedures are described in \S~2.  \S~\ref{res} reports the results of the multi-epoch study. In \S~\ref{srci}, we discuss  the morphological evolution of the radio continuum from Source~I and its interpretation in terms of an ionized disk. In \S~\ref{pas} we suggest that Source~I and BN had a past close passage, based on proper motion measurements. In \S~\ref{dyn}, we discuss dynamical scenarios that can explain the origin of the proper motions measured for  Source~I and BN. \S~7 and 8 discuss problems related to the mass of Source~I and a possible origin for the fast, wide-angle H$_2$ outflow, respectively.  Finally, conclusions are drawn in \S~9.

%-----------------------------------------------------------------------------
 % TABLE 1
%\begin{deluxetable}{ccccccc}
\begin{deluxetable*}{lllllll}
\tabletypesize{\scriptsize}
%\tabletypesize{\footnotesize}
\tablewidth{0pc}
%\tablenum{1}
\tablecaption{Parameters of Observations.}
\tablehead{
\colhead{Program} & \colhead{Date} & \colhead{$T$}\tablenotemark{a}  & \colhead{Synthesized Beam ($R=0$)}\tablenotemark{b} & \colhead{RMS} & \colhead{Synthesized Beam ($R=8$)}\tablenotemark{b}  & \colhead{RMS}\\ 
\colhead{} & \colhead{} & \colhead{(h)}  & \colhead{$\theta_M('') \times \theta_m(''); \ P.A.(^{\circ})$} & \colhead{(mJy/bm)} & \colhead{$\theta_M('') \times \theta_m(''); \ P.A.(^{\circ})$} & \colhead{(mJy/bm)}
} 
\startdata
%\hline
AM668A\tablenotemark{c} & 2000/11/10  & 6.6 & 0.041 $\times$ 0.028; -30 & 0.14 & $0.058 \times 0.045;\ -20$ & 0.13  \\ 
AG622 \tablenotemark{d} & 2002/03/31 &2.7 & 0.046 $\times$ 0.034; -4 & 0.16 & $0.066 \times 0.047;\ -17$  & 0.14 \\
AC817 & 2006/04/15 & 0.9& 0.051 $\times$ 0.030; 11 & 0.28& $0.066 \times 0.044;\ -5$  & 0.24  \\
AC952 & 2009/01/12 & 3.3& 0.058 $\times$ 0.039; 3 & 0.11& $0.079 \times 0.053;\ +4$  & 0.11  \\
%\hline
\enddata
\tablecomments{\\
%COMMENT FROM LDM: ADD AN ``EXPLANATION OF COLUMNS'' LIST HERE
\scriptsize
\tablenotemark{a}{Approximate on-source integration time.} \\
\tablenotemark{b}{Synthesized beams correspond to images made using the AIPS task IMAGR with a robust parameter $R=0$ and $R=8$, respectively.}\\
\tablenotemark{c}{Archival data from the program AM668A have been published by \citet{Rei07}}. \\
\tablenotemark{d}{Data from the program AG622 have been published by \citet{Rod05}.}
}
\label{obs}
%\end{deluxetable}
\end{deluxetable*}
%-----------------------------------------------------------------------------

%-----------------------------------------------------------------------------
\begin{deluxetable*}{cccccccc}
%\begin{deluxetable}{cccccccc}
\tabletypesize{\scriptsize}
%\tabletypesize{\footnotesize}
\tablewidth{0pc}
\tablecaption{Parameters of the 7~mm sources in Orion BN/KL}
\tablehead{
\colhead{Sources} & & \colhead{$\alpha$ (J2000)} & \colhead{$\delta$ (J2000)} & $e_t \ e_n \ e_s \ e_{\nu}$  &\colhead{Flux Density} &\colhead{Angular Size} \\ 
\colhead{} & & \colhead{($^{h\ m\ s}$)} & \colhead{($^{\circ} \ ' \ ''$)}  & (mas) &\colhead{(mJy)} &\colhead{$\theta_M('') \times \theta_m(''); \ P.A.(^{\circ})$}}
\startdata
BN & &   05 35 14.1094 & -05 22 22.724 & 5 0.3 1 $\lesssim 2$ & $23\pm2$  & $0.095 \pm 0.007 \times 0.072 \pm 0.008; 38 \pm 10 $\\ 
I & &   05 35 14.5141 & -05 22 30.575 & 5 1 1  $\lesssim 2$ & $11\pm2$  & $0.23 \pm 0.01 \times 0.12 \pm 0.01; 142 \pm 2$\\ 
n & &   05 35 14.3571 & -05 22 32.719 & 5 4 0  $\lesssim 2$ & $1.2\pm0.1$ & $0.126 \pm 0.018 \times 0.064 \pm 0.01; 6 \pm 8$\\ 
H & &   05 35 14.5008 &  -05 22 38.691 & 5 2 0  $\lesssim 2$ & $1.3\pm0.1$  &$0.075 \pm 0.015 \times 0.053 \pm 0.01; 4 \pm 8$\\ 
\enddata
\tablecomments{
%\footnotesize
\scriptsize 
The source names refer to  \citet{Lon82}, \citet{Gar87}, and \citet{Chu87}. 
 The positions, flux densities, and sizes reported here are from the
 January 12 2009 epoch. Angular sizes were derived as described in the Text.
 The quoted uncertainties represent the dispersion among images with various weighting schemes. Individual contributions to the total error budget ($\sim 5$~mas) on absolute positions are listed: $e_t, \ e_n, \ e_s, \ e_{\nu}$ (see Appendix).
%Absolute positional accuracy is estimated to be $\sim 5$~mas. 
}
\label{sources}
\end{deluxetable*}
%\end{deluxetable}
%===========================
\section{Observations and Data Reduction}
%===========================
\label{obser}
Observations of the $\lambda7$~mm continuum emission in Orion BN/KL were made
using the Very Large Array (VLA) of the National Radio Astronomy
Observatory (NRAO)\footnote{NRAO is a facility of the National Science Foundation
operated under cooperative agreement by Associated Universities, Inc.}.
We present data obtained in four distinct epochs spanning over 8 years
(see Table~1).  All data were obtained while the VLA was in the
A-configuration, yielding a resolution of approximately 0\pas05\footnote{Project AM668A was conducted in A configuration plus the Pie Town VLBA antenna.}.
%Table~\ref{obs} summarizes the observational parameters of the different programs.  
%The total time on source varied in the range x1-x2hr for different epochs, usually interleaved with other observations for more complete u-v coverage. 

In order to image the continuum emission from radio sources in BN/KL
and to provide a strong signal as a phase-reference to calibrate the
phase and amplitude of the (weak) continuum signal, we employed a
dual-band continuum setup with a narrow band (6.25~MHz in 
programs AC952 and AC817; 3.25~MHz in AG622; and 1.56~MHz in AM668A), 
centered on the SiO $v = 1, J = 1-0$ line (rest frequency 43122~MHz),
and a broad band 
(50 MHz) centered on a line-free portion of the spectrum (offset in
frequency from 
the maser by: $\sim$350~MHz for AC952, $\sim$100~MHz for AC817 and AG622, $\sim$40~MHz for AM668). Both frequency bands were observed in dual-circular polarizations. 
Absolute flux density calibration was obtained from observations of 3C286 or 3C48 (depending on the epoch). Short (60~s) scans of the nearby (1.3$^{\circ}$) QSO J0541$-$0541 (measured flux density in the range 0.7-1.5~Jy, depending on the epoch) were alternated with 60~s scans of Source~I to monitor amplitude gains, tropospheric and instrumental phase variations, and to determine the electronic phase offsets between the bands.\footnote{In AM668A, the QSO 0501$-$019 was observed every 25 minutes (relatively infrequently compared with the timescale of tropospheric and instrumental phase variations), so no absolute astrometry was available for that program.} % The continuum in AG622 does not have FS, absolute astrometry comes from the maser obs (in FS).
We then self-calibrated the narrow-band $v=1$ SiO maser data, 
applied the derived phase and amplitude corrections 
to the broadband continuum data, and produced a high-quality map of
the continuum emission. A detailed description of this calibration procedure can be found in \citet{God09}. 
 
  Using the Astronomical Image Processing System (AIPS) task IMAGR, we
  made images of the BN/KL region with 4096$\times$4096 pixels and cell size 0\pas005 to cover a 20$''$ field.%, which includes BN, Source~I, and Source~n.
 We produced maps with two different weightings of the (u,v)
 data,  first setting the IMAGR ``ROBUST'' weighting parameter to $R=
 8$ (equivalent to natural weighting) and then to $R$=0.  
We restored these image with a circular beams with FWHM 59~mas and
47~mas, respectively. The sizes of these restoring beams are
approximately the average of the geometric-mean sizes among different epochs (see Table~\ref{obs}). Since the synthesized beam size in the AC952 program was larger  ($\sim$20\%) than that in other programs,  this choice resulted in a degradation of the resolution for other epochs. For this reason, we also produced images with round restoring beams of 40~mas ($R=0$ weighting) and 50~mas ($R= 8$ weighting), equal to the geometric-mean size of the beams in projects AC817 and AG622, thus slightly ``super-resolving'' AC952. The ``high-resolution" maps were used to analyze structural changes in the continuum  emission from Source~I (Sect.~\ref{srci}), whilst   the ``low-resolution" images were used to fit positions and track proper motions of the continuum sources in BN/KL (Sect.~\ref{pas}), since they minimize structural effects in locating the peak of emission. 
 
%===========================
\section{Results}
%===========================
\label{res}

Four continuum sources were detected above a $6\sigma$ lower limit of $\sim0.1$~mJy~beam$^{-1}$ in the most sensitive epoch (2009 January 12): BN, I, {\it n}, H \citep{Lon82,Gar87,Chu87}. 
For each detected source, a two-dimensional ellipsoidal Gaussian model
was fitted in a $100 \times 100$ pixel area around the peak
emission, using the AIPS task JMFIT. For Source~I, we used the AIPS
task OMFIT to directly fit the visibilities using a disk model
geometry, which is more appropriate for the elongated morphology of 
the emission than the Gaussian model fitted by
JMFIT (see below). 
The positions, flux densities, and deconvolved angular sizes of the detected sources for the 2009 January 12 epoch are given in Table~\ref{sources}. 
%Positions and sizes are given by fits with JMFIT. Integrated fluxes are measured with the TVSTAT task, by integrating all the flux present within a rectangular window well fitted to each source emission.
The flux densities obtained for BN and Source~I are in good agreement with the values previously reported in the literature \citep{Men95,Rei07}. 
source~{\it n} was recently detected with the VLA for the first time at 7~mm by \citet{Rod09}, who reports a flux density of $2\pm0.2$~mJy. This a factor $\sim$2 higher than what we found in 2009 January 12. The  discrepancy  might be due either to a different filtering of the total flux in the two observations (0\pas05 vs 0\pas2 beam) or to intrinsic variability of  source~{\it n}, possibly due to non-thermal nature of the continuum emission \citep{Men95}. Based on a comparison between 7~mm and  3.6~cm observations, 
 however, \citet{Rod09} estimated a spectral index of 0.2 for the
 continuum emission from source~{\it n}, suggestive of marginally
 thick free-free thermal emission, as expected in an ionized
 outflow. Since the 3.6~cm emission is extended over 0\pas6 (i.e., an
 order of magnitude more than the restoring beam in A-configuration),
 the variation in flux between measurements with different angular
 resolutions is most probably due to filtering of the total flux in the most extended configuration.

\begin{figure}
\centering
\includegraphics[width=0.5\textwidth,trim=2cm 1cm 3cm 0cm,clip]{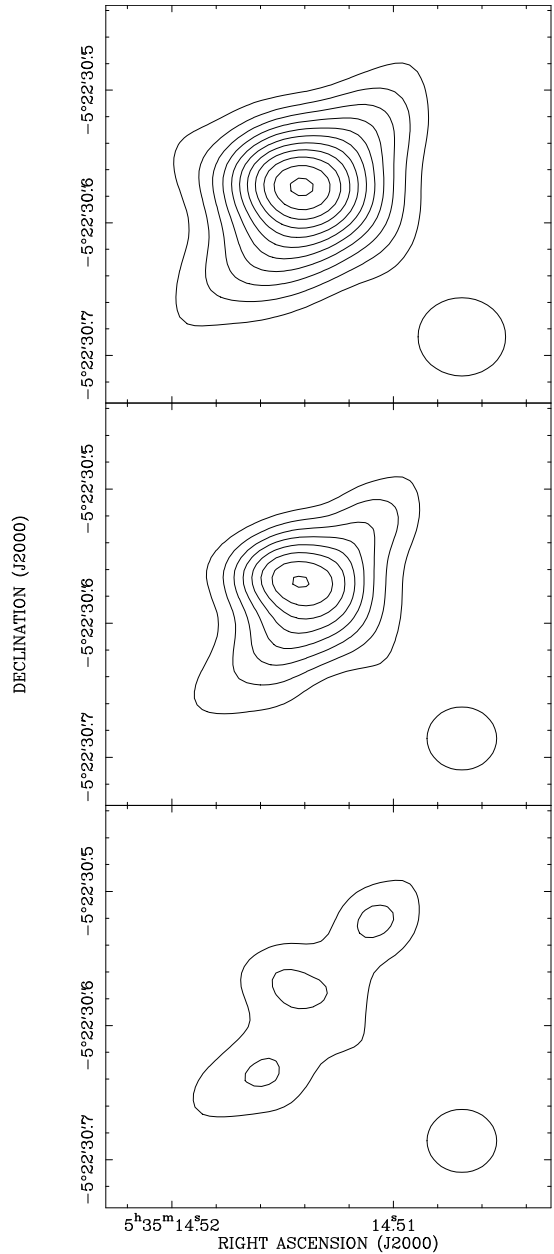}
\caption{Continuum images of Source I in Orion BN/KL at 43~GHz made with the VLA in the A configuration (2009 January 12). The images have been produced with natural weighting and a round restoring beam of 59~mas ({\it upper panel}) and with a $R=0$ weighting and a round restoring beam of 47~mas ({\it middle and lower panel}), respectively. 
In order to demonstrate the elongated structure, we subtracted a  (6~mJy, 50~mas size) Gaussian component at the centroid of the emission ({\it lower panel}). In all images the contour levels are at integer multiples of 0.4~mJy~beam$^{-1}$. The FWHM of the restoring beams are shown in the bottom right corner of each panel. 
We interpret the northwest-southeast elongated emission, as coming
from an ionized disk surrounding Source~I \citep[see also][]{Rei07}.%At a distance of 414 pc, 0\pas1 corresponds to 40~AU. 
%COMMENT BY LDM: ARE THE BEAMS SHOWN HERE ACTUALLY CIRCULAR? THEY LOOK SLIGHTLY ELLIPTICAL
}
\label{srci_ima}
\end{figure}

% SRCI
Our images of the continuum emission from Source~I at 43 GHz for the
2009 January~12 epoch are shown in Figure~\ref{srci_ima}, where different weightings and restoring beams were used (59~mas, top panel; 47~mas, middle panel). 
The emission is composed of a compact component at the center and a component elongated NW-SE. 
In order to better demonstrate 
the structure of the elongated component, we have subtracted a 6~mJy (beam-sized)  Gaussian component at the centroid of the emission. The resulting image (Fig.~\ref{srci_ima}, bottom panel) reveals a structure with a radius of $\sim 45$~AU and a brightness of about 4.5~mJy~beam$^{-1}$, elongated at a position angle (P.A.) of $142\pm2^{\circ}$. It is barely resolved along the minor axis and has a deconvolved axial ratio of about 2:1. 
The radius of this elongated structure agrees to within uncertainties
with previous measurements reported by \citet{Rei07}. \citet{Rod09} report a smaller size for Source~I at 7~mm (0\farcs09$\pm$0\farcs01$ \times $0\farcs06$\pm$0\farcs02).  Their image, however, has lower sensitivity than the ones reported here, owing to shorter integration time and narrow bandwidth employed (line-free channels were used to map the continuum), hence the quoted size refers to the central compact component while the elongated component is not detected. 

\begin{figure}
\centering
\includegraphics[width=0.5\textwidth]{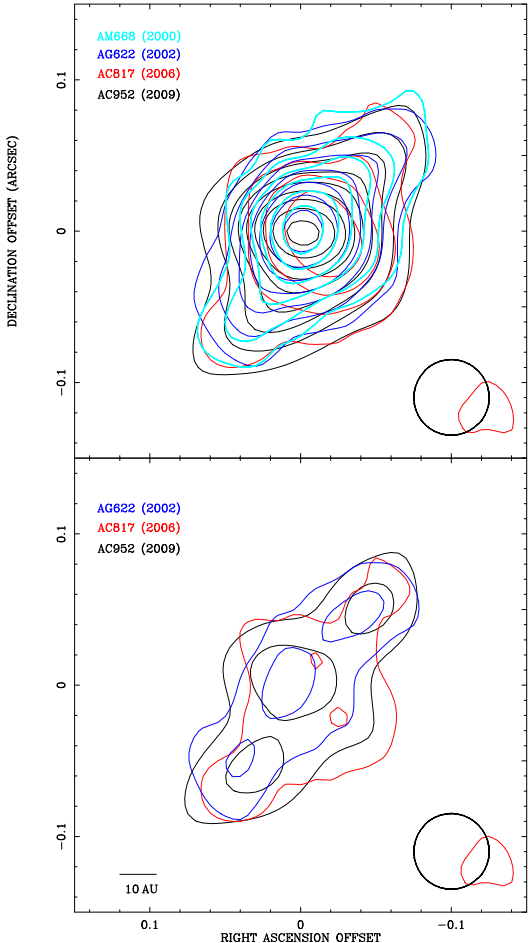}
\caption{Continuum images of Source I in Orion BN/KL at 43~GHz
  observed with the VLA in the A configuration at several epochs ({\it
    upper panel}). We subtracted the brightest component near the
  center  in each epoch ({\it lower panel}). %In order to facilitate the comparison, maps were produced with an average of the synthesized beam of all epochs.
 The images have been produced with natural weighting and a circular restoring beam of 50~mas ({\it bottom right corner}). 
The contour levels are at integer multiples of 0.4~mJy~beam$^{-1}$ (2000), 0.5~mJy~beam$^{-1}$ (2002), 0.7~mJy~beam$^{-1}$ (2006), 0.4~mJy~beam$^{-1}$ (2009). The choice of contours is dictated by differing sensitivity and uv-coverage in different experiments (Table~\ref{obs}).
%The FWHM of the restoring beams are shown in the bottom right corner of each panel. 
There is no significant change in the morphology or position angle of the structure over $\sim$8 years ($142^{\circ}\pm2^{\circ}$). 
}
\label{struct}
\end{figure}

Figure~\ref{struct} shows a comparison among images of Source~I at different epochs. Both size and position angle of the continuum emission appear to be remarkably constant over the last decade.

\begin{figure}
\includegraphics*[width=7cm,angle=-90,trim=0cm 0cm 1cm 0cm]{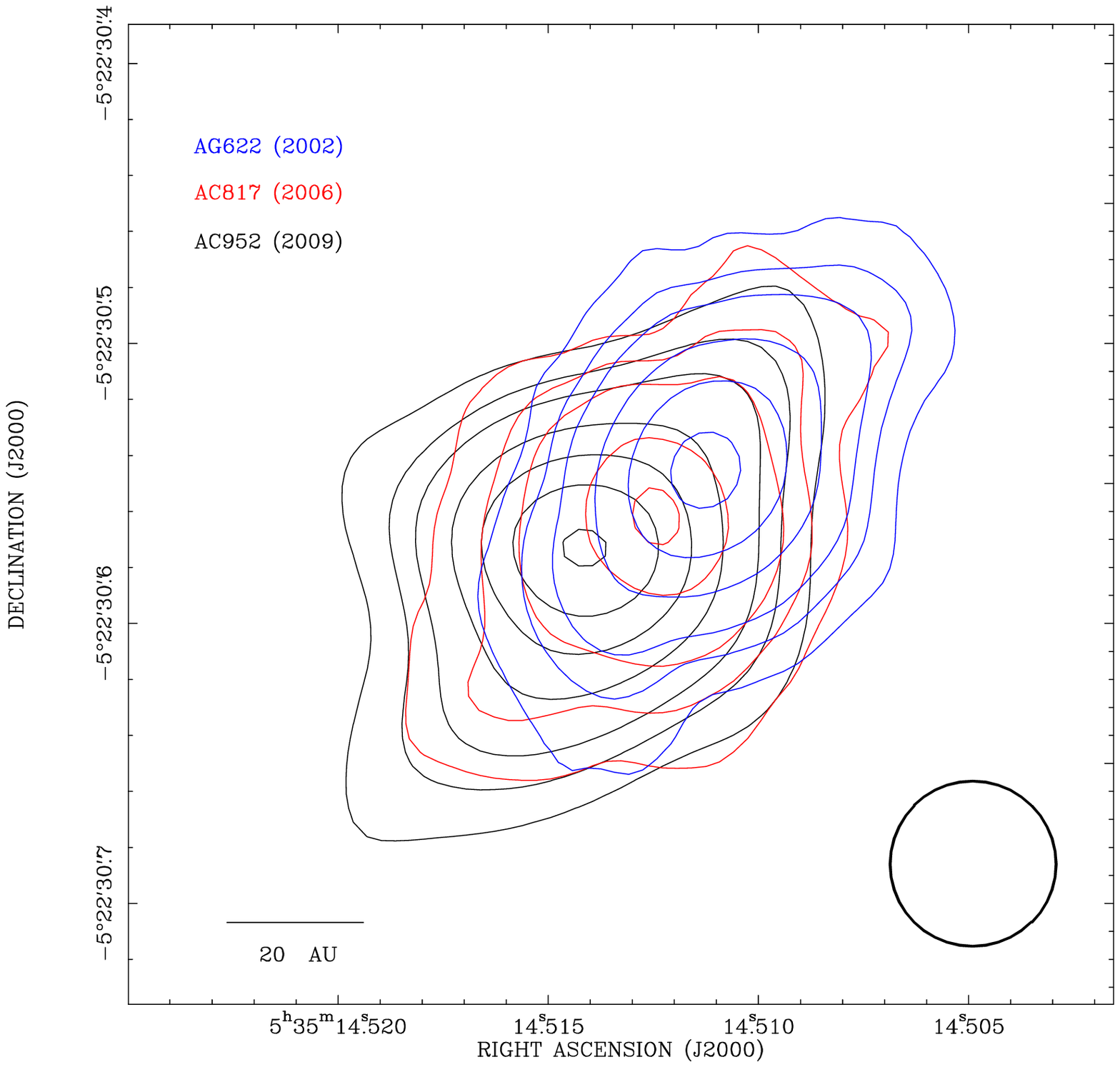}
\includegraphics*[width=7cm,angle=-90,trim=1cm 0cm 0cm 0cm]{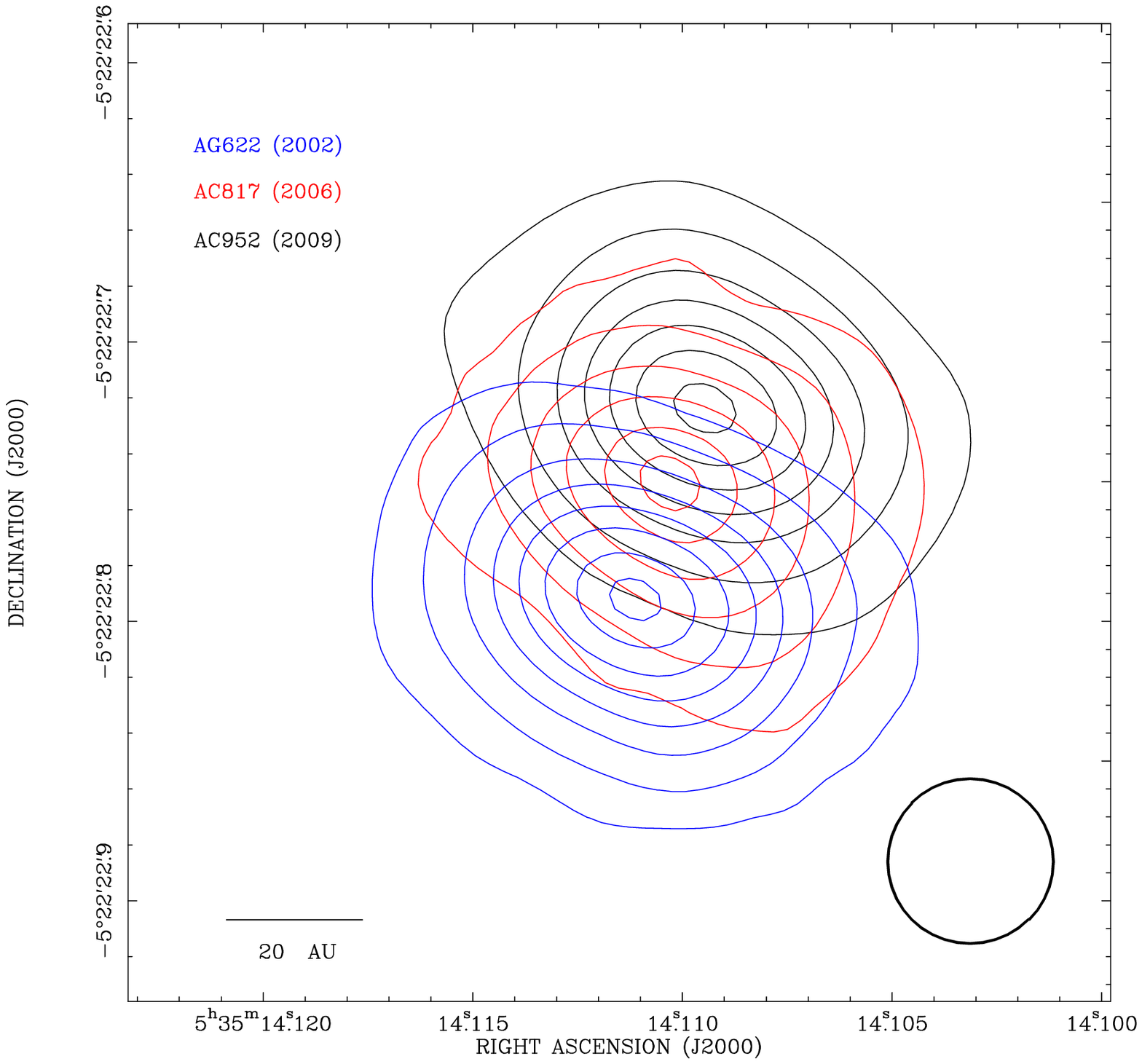}
\caption{Contours for the 2002 ({\it blue}), 2006 ({\it red}), and
  2009 ({\it black}) epochs of radio source I ({\it upper panel}) and the BN object
  ({\it lower panel}). Contours are (1, 2, 3, 5, 7, 9, 11) $\times$
  0.5 (2002), 0.7 (2006) and 0.4 (2009)~mJy~beam$^{-1}$ ({\it upper
    panel}), and (1, 3, 6, 9, 12, 15, 18, 20)  $\times$ 0.7 (2002),
  0.9 (2006) and 0.7 (2009)~mJy~beam$^{-1}$  ({\it lower panel}). The
  images have been restored with a circular beam with FWHM 59~mas 
(bottom right corner). The displacement of the peaks in each image
shows the 
absolute proper motions of the radio continuum emission of Source I and BN.}
\label{cont}
\end{figure}

\begin{figure}
\includegraphics*[width=4.2cm, trim=0cm 0cm 0cm 0cm,clip]{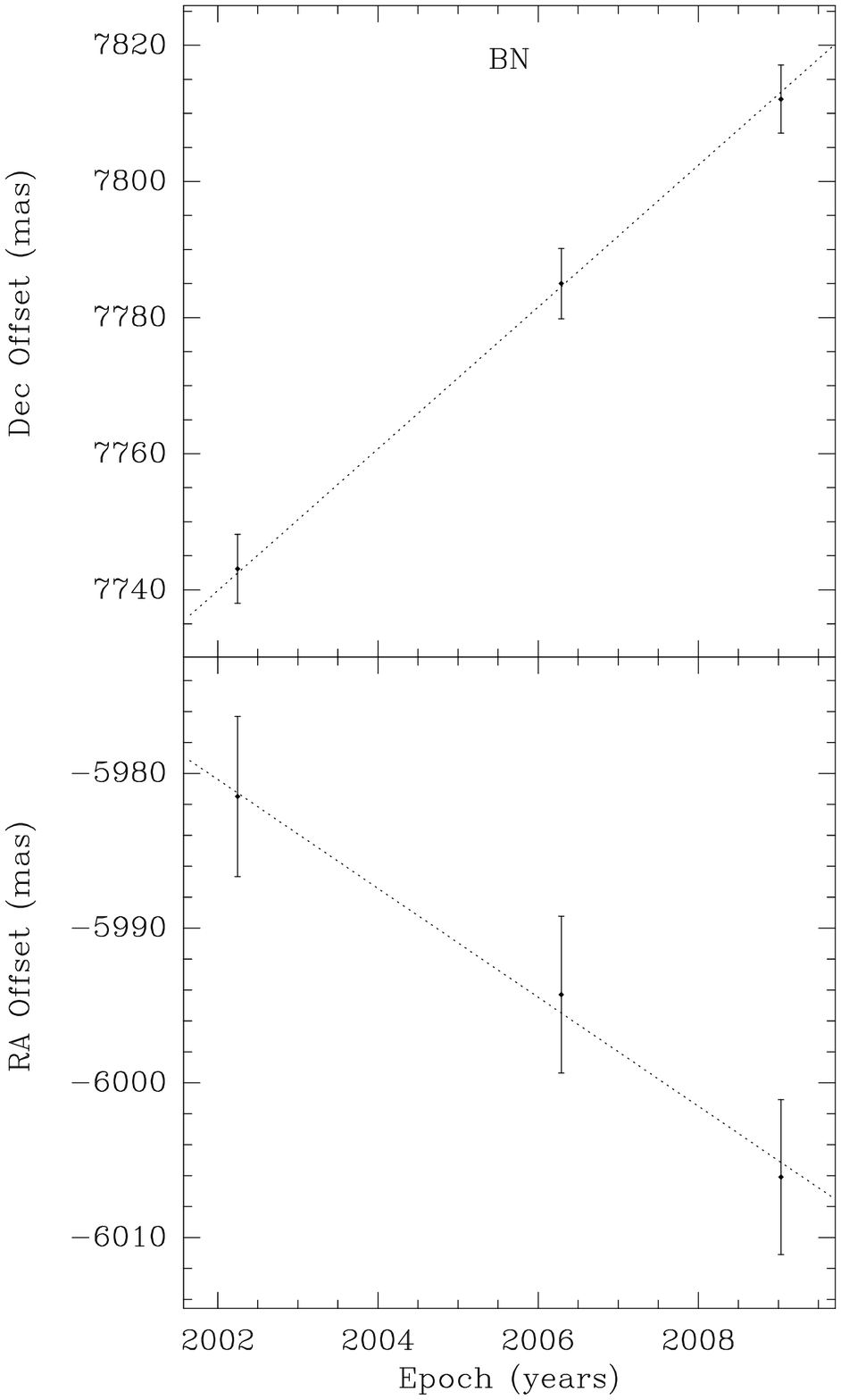}
\includegraphics*[width=4.2cm, trim=0cm 0cm 0cm 0cm,clip]{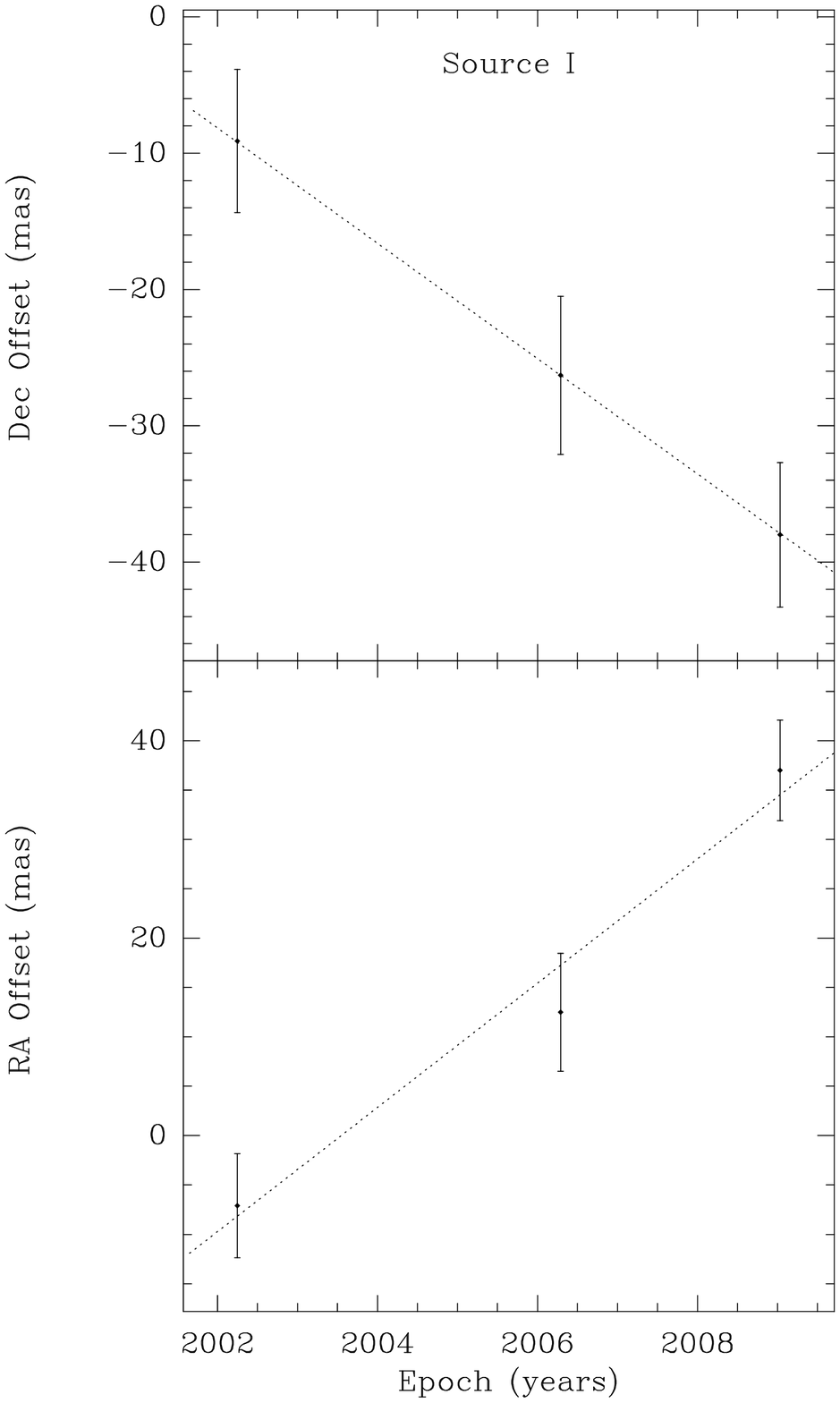}
\caption{Measured absolute proper motions in RA and DEC of the radio continuum centroid of Source I and BN,
 relative to position $\alpha(J2000) = 05^h 35^m 14^s.5116$,   $\delta(J2000) = -05^{\circ} 22' 30$\pas536. In each panel, the dotted line shows the proper motion calculated by the (variance-weighted) linear least-squares fit of the positional offsets with time.}
\label{fits}
\end{figure}

%pm determination+error budget
Extreme care was used in computing the absolute astrometry for the 
three new (non-archival) 
datasets reported here, the accuracy of which we quantify as
$\sim$5~mas (see Appendix). 
 This accurate absolute astrometry enabled us to track proper motions for BN and Source~I over three epochs. 
%Source {\it n} and H were detected at two epochs and only tentative proper motions could be derived (the results are not discussed here).
 Figure~\ref{cont} shows the contour images of BN and Source~I at
 different epochs. The displacement of the peaks in each image shows
 the absolute proper motions of the radio continuum emission for both
 sources. The consistency of morphology in the continuum emission at
 different epochs reinforces the robustness of the motion measurements.
 The proper motions have been calculated by performing an error-weighted linear least-squares fit of positions with time (Figure~\ref{fits}). % Fig.~\ref{fits} shows the time variation of the R.A. and DEC offsets and the best linear fit giving the proper motion.  
  The  uncertainties of the absolute proper motions are the formal errors of the linear least-squares fits and depend mainly on the error of the absolute position of individual sources at different epochs. The derived absolute proper motions and associated uncertainties are reported in  Table~\ref{pmt}.
 The measured proper motions are in the range 15-22~\kms, confirming the large proper motions of sources I and BN found previously at 8.4~GHz by \citet{Gom08}. 
In  Table~\ref{pmt}, we also report the absolute proper motions of both sources in the Orion Nebular Cluster (ONC) rest-frame and relative proper motion of BN with respect to Source~I. The ONC rest-frame is obtained by subtracting the mean absolute proper motions of 35 radio sources located within a radius of 0.1~pc from the core of the ONC, $\mu _{\alpha }\mathrm{cos}\,\delta =+0.8\pm 0.2$~mas~yr$^{-1}$; $\mu_{\delta }=-2.3\pm 0.2$~mas~yr$^{-1}$, PA=$341\pm5^{\circ}$ \citep{Gom05}. 

 Based on only two epochs, source~{\it n} and H do not show significant proper motions in right ascension or declination at a typical 3$\sigma$ upper limit of $3-4$~mas~yr$^{-1}$.

 %Source~n at 8.4 and 43GHz
 %Source~n at 8.4 GHz shows a peculiar double radio source  with a jet-like structure elongated north-south across 0\pas6. We detected a weak point source at 7~mm in the more sensitive 2002 and 2009. The extended nature of the emission at 8.4 GHz makes difficult the comparison with images/positions at 43~GHz.The position of n given in Table 1 is inconsistent with the extrapolation of the proper motions of this source discussed by gomez08. One should keep in mind the ambiguity in associating the position at different frequencies (N,S,c). The tentative estimation based on two epochs is consistent with very small proper motion (Xkms).  

%
%--------------------------------------------------------------------------------------
%--------------------------------------------------------------------------------------

%--------------------------------------------------------------------------------------
%--------------------------------------------------------------------------------------

%--------------------------------------------------------------------------------------
\section{Morphological evolution of the radio continuum: An ionized disk around Source I}
%\section{An ionized disk around Source I}
\label{srci}
The continuum emission from Source I shows an elongated structure with a position angle of $142^{\circ}\pm3^{\circ}$ and a size $\sim$45~AU (Fig.~\ref{srci_ima}). The uncertainty quoted here for the position angle represents the dispersion of values measured in several epochs, which is not significantly larger than the uncertainty in a single epoch  ($2^{\circ}$) reported in Table~\ref{sources}. 
 A comparison among images at different epochs does not show any evidence of changes in structure, size or position angle over 8 years (Fig.~\ref{struct}).
In addition, we found no evidence for significant variations in the total 7~mm continuum flux density of Source~I over the last decade, with all values being consistent with $11 \pm 2$~mJy. %[Here the scatter in flux density represents uncertainty rather than true variations among epochs.]

What is the nature of this ionized emission? 
The centimeter-to-millimeter wavelength spectrum of the entire source can be characterized as a power law with flux density $S_{\nu} \propto \nu^{1.6}$ \citep{Beu06}. The emission is optically thick up to 100~GHz, with a dust component dominating above 300 GHz \citep{Beu06}.
The radio emission has been interpreted either as an ionized jet with a NW-SE axis \citep{Bal05,Tan08a} or as an ionized disk with a NE-SW axis \citep{Rei07}. 
In the following, we will discuss three alternative interpretations of the radio continuum emission from Source I: 1) an ionized jet, 2) an equatorial wind, and 3) an accreting disk. We will show that only the latter is consistent with our multi-epoch data.
%a further case: a "static" ionized outflow cavity, as JT proposed

 If the emission originates in an ionized jet, one would expect to detect structural variations  owing to motions of individual jet components. In fact, when observed with high-angular resolution, radio jets do not show a smooth structure, but contain clumps. Using multi-epoch VLA observations, detection of proper motions as large as  $\sim$100-500~\kms have been reported in the components of radio continuum jets associated with both  low-mass YSOs (e.g., HH1-2 -\citealt{Rod00}) and
 high-mass YSOs  (e.g., Cepheus-A - \citealt{Cur06}; HH80-81 - \citealt{Mar95}). 
 % Multiepoch studies by Martí et al. (1998) reveal proper motions of clumps in the jet, which correspond to tangential velocities of at least 500~\kms.  Another radio jet is that in Cep A2 (Rodríguez et al. 1994), where proper-motion studies yield tangential velocities of 600 km s-1 (Curiel et al. 2006). 
 Assuming 20~\kms as a lower limit for the expansion velocity of the jet associated with Source~I (as estimated from the SiO maser flow by \citealt{Mat10}), displacements of $\sim$40~AU are expected on a temporal baseline of 9 years.  Such displacements are comparable to the radius of the continuum emission and are easily detectable with the linear resolution of our observations ($\sim$25~AU).  In principle, no structural changes would be expected from an optically thin, homogeneous ionized region. However, even in the case of uniform density, thermal pressure would drive expansion of the ionized region as a whole, as seen for ionized winds and HII regions (e.g., \citealt{Aco98}). As well,  classic models of spherical winds \citep{Pan75} and ionized jets \citep{Rey86} predict the size of the ionized region to vary as a function of frequency as $\nu^{-0.7}$, which would imply a factor of 3 change in size between 8.4~GHz and 43~GHz. Nonetheless, deconvolved sizes obtained for the radio continuum emission around Source~I are similar at the two frequencies: 0\pas23 $\times$ 0\pas115 at 43~GHz (this work) and 0\pas19 $\times <$0\pas15 at 8.4~GHz \citep{Gom08}.

 An alternative interpretation is that the radio emission from Source I arises in a disk-like structure rather than in a jet, where the gas is not accreting but rather consists of an equatorial ionized wind. Such an equatorial wind has been modeled by \citet{Dre98}. The model predicts that the radiation pressure (mediated by line opacity) from a massive YSO accelerates material from the surface of the disk and blows it away mainly in the equatorial plane. The velocities obtained for the gas are of the order of a few hundreds \kms. Also in this scenario, large changes in morphology are expected over short time intervals (i.e., a few years) owing to fast motions of clumps in the wind, as observed in the luminous YSO S140 IRS1 \citep{Hoa06}. In addition to fast motions, rapid changes in the wind structure are expected as well, owing to small changes in illumination by the ionizing YSO and to the  inherently unstable nature of the radiation-driven winds (Owocki et al. 1988). Hence, we conclude that our multi-epoch data are not consistent with the picture of a radiation-driven disk-wind model.

Finally, we consider the case  that the elongated structure traces ionized accreting gas in a disk. 
  \citet{Ket03} has shown that, even when the inner portion of a disk is fully ionized, accretion of (ionized and neutral) material can continue onto a massive YSO. In this spherically symmetric model, the gravitational attraction of the star  exceeds the outward force of the (radiation and thermal) pressure inside a critical radius $r_c = GM/2c_s$, where G is the gravitational constant, M is the mass of the YSO, and $c_s$ is the sound speed in the ionized material. 
 In the hypothesis of spherical accretion, by scaling the stellar parameters in Table~1 from \citet{Ket03} to a 10~\ms~star, ionized accretion can proceed inside of the critical radius $\sim$25~AU with an accretion rate $\sim 10^{-5}$~\msyr.  This critical radius is reasonably consistent with the disk size observed at 43~GHz in Source I. In Sect.~\ref{dyn}, we report evidence for Source~I being a binary of total mass of the order of 20~\ms. In the most probable case of an equal-mass binary, this would slightly increase the critical radius above, estimated in the assumption of one 10~\ms~single star.
  We note that similar values for the critical radius and accretion rates are obtained for models of ionized accretion in the presence of significant angular momentum \citep{Ket07}, as required by the disk geometry in Source~I. 
We also note that  an accretion rate of $10^{-5}$~\msyr and a central stellar mass of 10 to 20~\ms~is consistent with a luminosity of a few times $10^4$~L$_{\odot}$ for Source~I. 

The interpretation of the ionized elongated structure as a disk versus
a jet is relevant to a key longstanding question regarding Orion BN/KL: what
are the drivers of the region's outflows?
In our interpretation,  the elongated radio continuum structure traces
a nearly edge-on disk, with a spin axis aligned NE-SW, parallel to the
low-velocity outflow (see \S~\ref{int}). This is consistent
with the model for Source~I based on high-resolution monitoring of
several SiO maser transitions (\citealt{Kim08,Mat10,Gre04a} and in prep). 
Greenhill et al. (in prep.)  estimated a {\em mass-loss rate} of the NE-SW
outflow to be $3 \times 10^{-6}$~\msyr, assuming a gas density of
$10^6$~cm$^{-3}$ (required for the excitation of $v=0$ SiO masers),
a typical velocity of 18~\kms for the outflow, and a radius of
100~AU. Based on the models of \citet{Ket07}, we estimate an {\em
accretion rate} $\sim10^{-5}$~\msyr. A ratio of O(0.1) for the mass-loss 
rate to the accretion rate is  typical of values 
inferred from observations of low-mass pre-main-sequence stars (e.g.,
\citealt{Bon96}) and of values predicted in MHD ejection models
\citep{Shu00,Kon00}.
This might indicate that Source~I is still in an active accretion
stage. We caution that the  mass-loss rate of the outflow from
Source~I estimated from the SiO masers is accurate only to within an
order of magnitude (Greenhill et al., in prep.).  Nevertheless, the
preponderance of evidence strongly 
supports a picture where the high-mass YSO Source~I is
powering a 
disk/outflow system along a NE-SW axis.

%-----------------------------------------------------------------------------
 % 
\begin{deluxetable*}{llllllll}
%\begin{deluxetable}{llllllll}
\tabletypesize{\footnotesize}
\tablewidth{0pc}
\tablecaption{Proper Motions of Radio Source I and the BN object}
\tablehead{
\colhead{Sources} & \colhead{$\mu_{\alpha }$} & \colhead{$\mu_{\delta }$} & \colhead{$\mu_{\mathrm{tot}\,}$} & \colhead{$V_{\rm x}$} & \colhead{$V_{\rm y}$} &\colhead{$V_{\rm mod}$} & P.A.\\ 
\colhead{} & \colhead{(mas~yr$^{-1}$)} & \colhead{(mas~yr$^{-1}$)} & \colhead{(mas~yr$^{-1}$)} & \colhead{(\kms)} & \colhead{(\kms)} &\colhead{(\kms)} & (deg)
}
\startdata
\multicolumn{7}{l}{\it Absolute proper motions}\\
I &   $6.3\pm1.1$ &  $-4.2\pm1.1$ &  $7.6\pm1.1$ &    $12.3\pm2.1$ &  $-8.3\pm2.1$  &  $14.9\pm2.1$ &  $123.9\pm8.2$ \\
BN &  $-3.5\pm1.1$ &  $10.4\pm1.1$ &  $11.0\pm1.1$  &  $-6.9\pm2.1$ &   $20.4\pm2.1$ &  $21.5\pm2.1$   &  $341.3\pm5.5$ \\
&&&&&&&\\
\multicolumn{7}{l}{\it Proper motions relative to I}\\
BN & $-10.7\pm0.3$ &  $14.2\pm0.3$ &   $17.8\pm0.3$ &    $-21.0\pm0.6$ &  $27.9\pm0.6$ &  $34.9\pm0.6$ &  $323.0\pm1.0$ \\
&&&&&&&\\
\multicolumn{7}{l}{\it Proper motions in the ONC rest-frame}\\
I &   $5.5\pm1.1$ &  $-1.9\pm1.1$ &  $5.8\pm1.1$ &    $10.8\pm2.1$ &  $-3.8\pm2.2$  &  $11.5\pm2.1$ &  $109.4\pm10.9$ \\
BN &  $-4.3\pm1.1$ &  $12.7\pm1.1$ &  $13.4\pm1.1$  &  $-8.5\pm2.1$ &   $25.0\pm2.1$ &  $26.4\pm2.1$   &  $341.2\pm4.6$ \\
\enddata
%\tablecomments{}
%\footnotesize
%\tablenotetext{a}{Proper motions relative to I.} %REQUIRES \tablenotemark{a}
%\tablenotetext{b}{Proper motions in the ONC rest-frame.}
\label{pmt}
\end{deluxetable*}
%\end{deluxetable}
%
%-----------------------------------------------------------------------------
%--------------------------------------------------------------------------------------
%--------------------------------------------------------------------------------------
%--------------------------------------------------------------------------------------
%--------------------------------------------------------------------------------------

\section{Proper motions: A close passage between Source~I and the BN object}
\label{pas}
\begin{figure}
\centering
\includegraphics[width=0.5\textwidth]{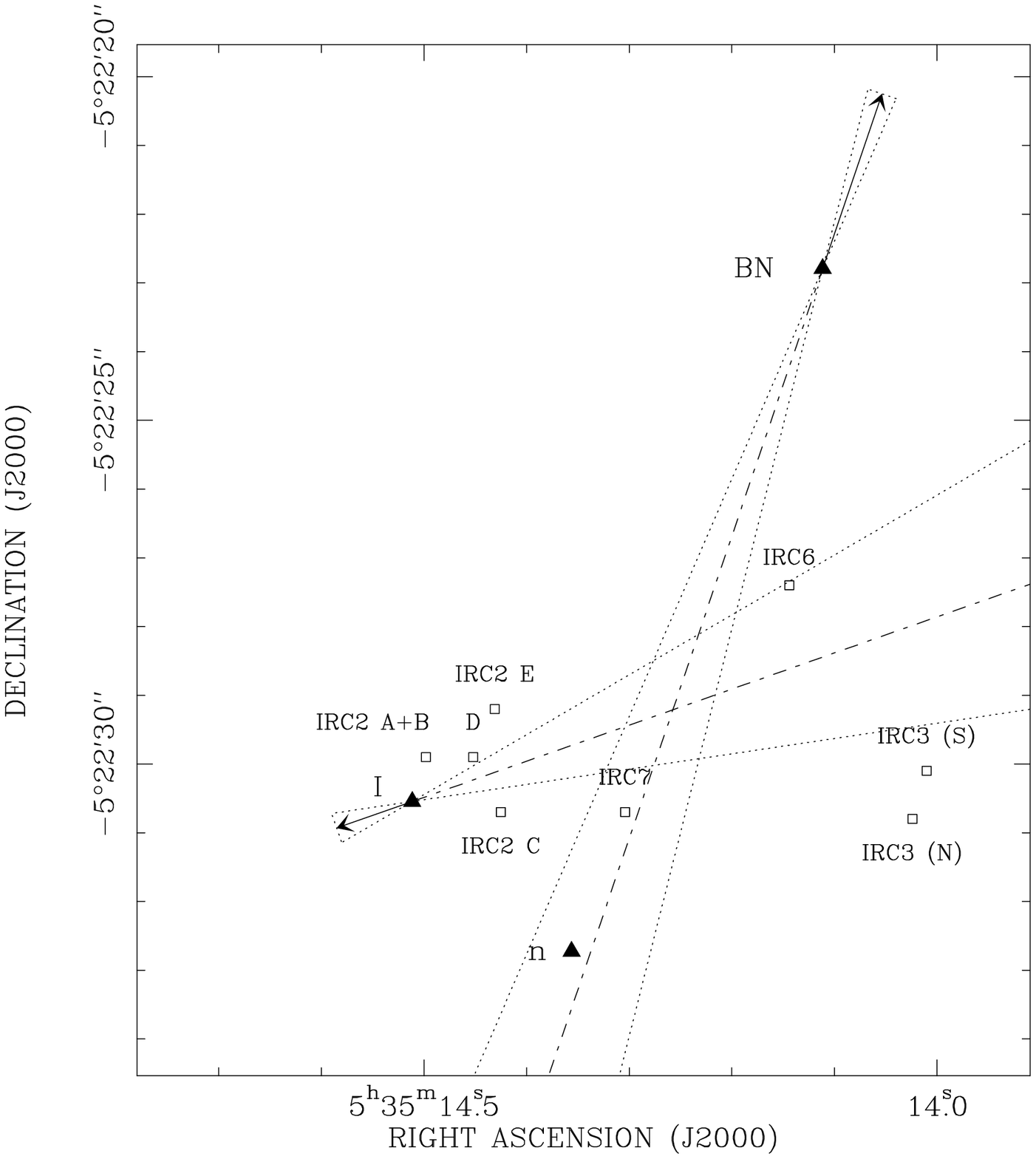}
\includegraphics[width=0.5\textwidth]{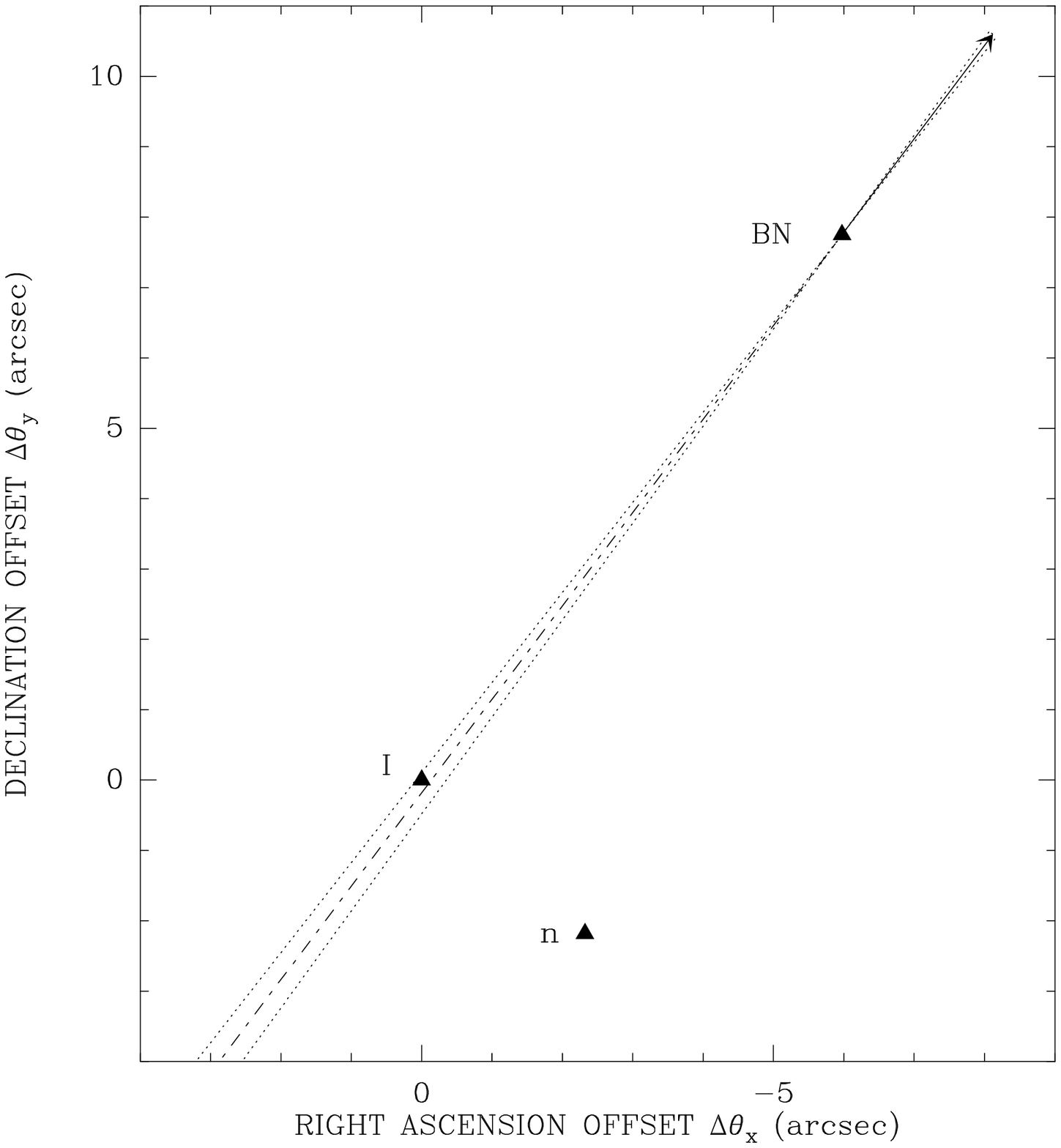}
\caption{Proper motions of BN and Source~I measured from our multi-epoch datasets in the ONC rest-frame ({\it upper panel}). Proper motion of BN relative to Source I ({\it lower panel}).
The arrows indicate proper motion direction and displacement for 200 yr. The dashed lines indicate errors in proper motion position angle, and they track backwards showing past positions of source. About 560 years ago, BN, Source~I, and several IR sources within the BN/KL region were located within a few arcseconds of each other. 
Our measurements are consistent with a null proper motion for source~{\it n}.
}
\label{pm}
\end{figure}

We observe BN and Source~I moving  with high speeds in opposite direction with respect to one another. This result confirms previous independent proper motion measurements at 8.4~GHz \citep{Gom08} and at multiple frequencies ranging from 8.4 to 43~GHz \citep{Rod05}. 
 Figure~\ref{pm} shows maps of the absolute proper motions of both sources in the ONC rest-frame (upper panel) and relative proper motion of BN with respect to Source~I (lower panel). %The ONC rest-frame is obtained by subtracting the mean absolute proper motions of 35 radio sources located within a radius of 0.1~pc from the core of the ONC, $\mu _{\alpha }\mathrm{cos}\,\delta =+0.8\pm 0.2$~mas~yr$^{-1}$; $\mu_{\delta }=-2.3\pm 0.2$~mas~yr$^{-1}$, PA=$341\pm5^{\circ}$ \citep{Gom05}. 
 %In the ONC-restframe the proper motions are: $\mu _{\alpha }\mathrm{cos}\,\delta =5.5\pm 1.1$~mas~yr$^{-1}$, $\mu_{\delta }=-1.9\pm 1.1$~mas~yr$^{-1}$, PA=$109\pm10^{\circ}$ (12~\kms) for Source~I; $\mu _{\alpha }\mathrm{cos}\,\delta =-4.3\pm 1.1$~mas~yr$^{-1}$ (), $\mu_{\delta }=12.7\pm 1.1$~mas~yr$^{-1}$, PA=$341\pm4^{\circ}$ (26~\kms) for BN.
 
 %Despite our measurements show that absolute proper motions of BN and Source~I (in the ONC-rest-frame)  deviate significantly from being aligned (Fig.~\ref{pm}, upper panel),  
The more accurate relative proper motions show that Source~I falls within the error cone of the BN past motion (Fig.~\ref{pm}, lower panel). 
Based on our accurate relative astrometry, we can estimate the minimum separation BN and Source~I had in the past in the plane of the sky by extrapolating the velocity vector backward in time and assuming no acceleration. Following \citet{Gom08}, the minimum separation is  
%\begin{equation}
$$S_{\rm min}(^{\prime \prime })=\frac{|x_{2002.25}\mu_y-y_{2002.25}\mu_x|}{\sqrt{\mu_x^2+\mu_y^2}}$$
%\end{equation}
which occurs at an epoch $t_{min}$ given by
%\begin{equation}
$$T_{\rm min}(\rm years)=2002.25-\frac{|x_{2002.25}\mu_x+y_{2002.25}\mu_y|}{\mu_x^2+\mu_y^2}$$
%\end{equation}
where both quantities are expressed as a function of the relative position for program AG622 ($x_{2002.25} =-5.974^{\prime \prime }\pm 0.002^{\prime \prime }$, $y_{2002.25} =7.752^{\prime \prime }\pm 0.002^{\prime \prime }$) and relative proper motions of BN with respect to Source~I 
 ($\mu_{x}=-0.0107^{\prime \prime }\pm 0.0003^{\prime \prime }  \rm yr^{-1}$, and $\mu_{y}=0.0142^{\prime \prime }\pm 0.0003^{\prime \prime } \rm yr^{-1}$). 
%The relative proper motions are consistent within error, to those reported by \citet{Gom08}. 
By substituting these values, we obtain: $S_{\rm min}=$0\pas109$\pm$0\pas175 and $T_{\rm min}=1453\pm9$ years. Hence, our new measurements indicate that, about 560 years ago, BN and Source~I were as close as $50\pm 100$~AU in projection on the plane of the sky. 

Although we do not have position information along the line-of-sight
that confirms the close passage in 3-dimensions (proper motion analysis gives only a 2-dimension description of the dynamics), the line of sight velocities of BN and Source~I are also consistent with a dynamical encounter. Based on the measured radial velocities of the ambient molecular cloud (\vlsr=9~\kms: \citealt{Gen89}), BN (\vlsr=20~\kms, or 11~\kms with respect to the ambient cloud: \citealt{Rod09}), and Source~I (\vlsr=5~\kms, or $-$4~\kms with respect to the ambient cloud: \citealt{Mat10}), the ratio of radial velocities between BN and Source~I is $\sim$2.7, which is consistent with the value of 2.3 determined from the (ONC-frame) velocities in the plane of the sky. Hence, 5 to 6 observables are consistent with the hypothesis of a close passage between Source~I and BN.
We also note that the closest approach separation of 50~AU is comparable to the radius of the ionized disk observed today around Source~I (Sect.~\ref{srci}). As this is compact for a high-mass YSO accretion disk, we suggest that it may have been  truncated in the close encounter with BN. We note also that the ionized envelope surrounding BN has a comparable radius of $\sim 40$~AU (Table~\ref{sources}), consistent with the truncation hypothesis. 

% ARGUMENTS AGAINST SOURCE n ROLE IN BN-I INT.: 1) P.MO(43GHZ)~0; 2) WRONG ORIENTATION; 3) P.MO(8.4GHZ) CONTROVERSIAL.
\citet{Gom08} measured a large proper motion for source~{\it n} as well ($\mu=13\pm1$~mas~yr$^{-1}$, $v=25\pm2$~\kms, P.A.$=180^{\circ}\pm4^{\circ}$) using position measurements at 8.4~GHz and proposed a close passage with BN and Source~I. This result however disagrees with our measurements. 
%We tend to exclude that source~{\it n} played any major role in the dynamical interaction between Source~I and BN.
Using two epochs in which we achieved $5-6\sigma$ detections at 7~mm wavelength, we measured a much smaller proper motion of $3.7\pm1.6$~mas~yr$^{-1}$ and PA$=26\pm22$ in the ONC rest-frame. This indicates that source~{\it n} is moving either very slowly or not at all (our measurement is consistent with 0 within $\sim 2\sigma=3.2$~mas~yr$^{-1}$). In addition, the proper motion vector points towards and not away from the interaction center, as it would be required if source~{\it n} were truly ejected by a close passage with Source~I and BN.
As a final point, the  proper motion of source~{\it n} at 8.4~GHz reported by \citet{Gom08} should be viewed with a degree of caution. The emission at 8.4~GHz from source~{\it n} is resolved into a bipolar or elongated structure, depending on epoch (size$\sim$0\pas6), and the systematic error in the motion may arise from  internal variability. Tracking of northern and southern components of source~{\it n} by \citet{Gom08} may be possible, but a longer time baseline is required for certainty.  Of some concern, the dataset from \citet{Gom08} is not homogeneous in terms of observing set-up among epochs, and hence changes in uv-coverage could mimic change in structure/position. Finally, in their Figure~2, \citet{Gom08} show that the displacement of  source~{\it n} in the plane of the sky from the cluster origin is approximately half that of BN.  However, this is inconsistent with the two objects having comparable proper motions and starting their outward motion from the same point of origin and epoch (assuming linearity and constancy of motion). Further observation of source~{\it n} is required to confirm the proper motion measured at 8.4~GHz and hence the dynamical interaction with BN and Source~I. %The 7~mm band is preferable because there the emission is compact and modeling errors are thereby reduced.

In the following section, we will reconstruct the dynamical history of BN/KL based on the proper motions measured for Source~I and BN. Section~\ref{quadruple} will discuss the possible dynamical role of source~{\it n}, in the assumption that the 8.4~GHz proper motion measurement of Source~{\it n} derived by \citet{Gom08} is correct.
 
 %--------------------------------------------------------------------------------------
%--------------------------------------------------------------------------------------

 %-------------------------------------------------------------
 \section{Dynamical history of BN/KL}
\label{dyn}
 Observations have increasingly shown that stars are almost never born
 in isolation but favor formation in clusters, where interactions can
 occur among members of a collapsing protostellar group. Interactions
 with sibling stars play an even more important role in the formation
 of high-mass stars \citep{Bal05}. YSOs are also surrounded by
 circumstellar disks and envelopes, which can play a role in this
 context by increasing the interaction radius among members of the
 collapsing protostellar cluster \citep[e.g.][]{Moe06}.
The anamolously large relative motions between BN and Source~I
suggest that interactions have also played an important role in
Orion~BN/KL.
In the following sections, we investigate four scenarios to account
for the observed dynamics of the region:
1) the motions are independent and BN is a runaway star from the Trapezium cluster (Sect~\ref{teta1}); 
2) a close passage between the two protostars resulted in a star-disk interaction (Sect~\ref{double});
3) the motions are the result of a multiple ($\ge 3$) body interaction within a cluster of individual YSOs, where a tight massive binary is formed (Source~I) that carries away the binding energy of the system and BN is ejected (Sect~\ref{multiple}); 
4) a close passage between BN and Source~I, a pre-existing massive binary, resulted in the acceleration of both stars and the hardening of the original binary (Sect~\ref{bin}).

%-------------------------------------------------------------
\subsection{Scenario 1: Dynamical ejection of BN from $\theta ^{1} $ Ori C}
%-------------------------------------------------------------
\label{teta1}
We consider first the possibility that the motions of BN and Source~I are independent. \citet{Tan04,Tan08b} proposed that BN was ejected from the Trapezium by dynamical interaction involving the $\theta ^{1} $ Ori C  binary system. 

In a triple system usually the lightest member is ejected, sometimes as a runaway star, while an eccentric binary is formed between the remaining stars. The formation of a tight binary is required if the system was originally bound,  as a consequence of energy conservation. Usually, both components (star and close binary) leave their parental envelope, with velocities inversely proportional to their masses, as a consequence of momentum conservation \citep[e.g.][]{Rei00}.

\begin{figure}
\centering
\includegraphics[width=0.5\textwidth]{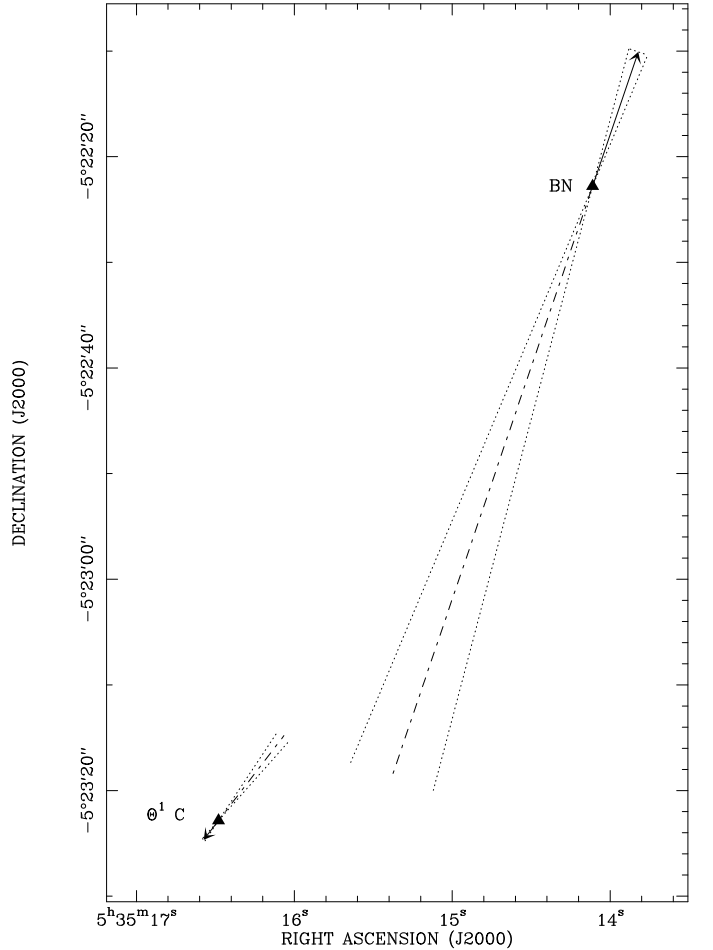}
\caption{Absolute proper motions of BN (radio, this work) and $\theta^{1}$ Ori C \citep[optical,][]{Vanalt88} measured in the ONC-rest-frame. % ({\it upper panel}). Proper motion of BN relative to $\theta^{1}$ Ori C ({\it lower panel}).
The arrows indicate proper motion direction and displacement on the plane of the sky over 1000 years. The dash-dot line indicates motion backward over 4400 years, assuming it is linear. The dashed lines indicate 1~$\sigma$ uncertainties in position angles. The minimum separation on the sky between BN and $\theta^{1}$ Ori C was $10.9^{\prime \prime } \pm 4.8^{\prime \prime }$.
}
\label{teta1cfig}
\end{figure}

$\theta^1$~Ori~C seems to have the physical properties predicted for such a dynamical event: an  eccentric, massive binary with primary and secondary masses greater than that of BN \citep{Kra07}; a total orbital energy ($E_{tot}\sim 3 \times 10^{47}$~erg) greater than the kinetic energy of BN ($10^{47}$~erg); a proper motion roughly in the opposite direction to the motion of BN and with approximately the predicted magnitude (for the dynamical mass of BN).
However, our absolute proper motion measurement is inconsistent with a close encounter between BN and $\theta^1$~Ori~C in the past. 
Figure~\ref{teta1cfig} shows the radio-ONC-frame\footnote{The radio-ONC-frame is defined by the mean absolute proper motions of 35 radio sources located within a radius of 0.1~pc from the core of the ONC, $\mu _{\alpha }\mathrm{cos}\,\delta =+0.8\pm 0.2$~mas~yr$^{-1}$; $\mu_{\delta }=-2.3\pm 0.2$~mas~yr$^{-1}$, PA=$341\pm5^{\circ}$ \citep{Gom05}.} proper motion of BN, $\mu_x=-4.3\pm 1.1\ \mathrm{mas}\,\ \mathrm{yr}\,^{-1}$, $\mu_y=+12.7\pm 1.1\ \mathrm{mas}\,\ \mathrm{yr}\,^{-1}$ (Table~\ref{pmt}), and the optical-ONC-frame\footnote{The optical-ONC-frame is defined by the mean absolute proper motions of 73 stars brighter than $V\sim12.5$ located within a radius of one-half degree of the Trapezium, corresponding to an average proper motion dispersion of $0.70\pm 0.06$~mas~yr$^{-1}$ \citep{Vanalt88}.} proper motion of $\theta^1$~Ori~C, $\mu_x =+1.4\pm 0.17 \ \mathrm{mas}\,\ \mathrm{yr}\,^{-1}; \mu_y=-1.8\pm 0.16 \ \mathrm{mas}\,\ \mathrm{yr}\,^{-1}$ \citep{Vanalt88}.
%We note that the proper motions determined by van Altena et al. (1988) are not in an absolute reference frame (their system is defined by the average proper motions of their reference stars), and the reliability of the comparison is uncertain. Tian et al. (1996) have discussed the different reference frames for the proper motions of stars in Orion obtained by five different groups and find that they differ at the ~1 mas yr-1 level. Furthermore, none of these reference frames is absolute in the sense of being referred to the remote quasars. 
The plot shows that past trajectories of BN and $\theta ^{1} $~Ori~C in fact intersect far from the Trapezium cluster. By repeating the same analysis we made for the relative motions of BN with respect to Source~I (this time referring positions and velocities to $\theta ^{1} $ Ori C\footnote{For $\theta ^{1} $ Ori C we took the absolute position from the COUP X-ray survey: $05^h35^m16.4789^s,\-05^{\circ}23'22.844''$, known to an accuracy of 0\pas3 \citep{Get05}.}), we obtain $x_{BN} =-35.4^{\prime \prime }\pm 0.3^{\prime \prime}$, $y_{BN} =60.0^{\prime \prime }\pm 0.3^{\prime \prime}$ , $\mu _{x}=-0.0057^{\prime \prime }\pm 0.0011^{\prime \prime }$~yr$^{-1}$, and $\mu _{y}=0.0145^{\prime \prime }\pm 0.0011^{\prime \prime }$~yr$^{-1}$ (P.A.=$339\pm4^{\circ}$) for the {\it relative} proper motions between BN and $\theta^1$~Ori~C (today). This yields $S_{\rm min}=10.9^{\prime \prime }\pm4.8^{\prime \prime }$ and $T_{\rm min}=-2403\pm312$ years.
  Hence, our measurements indicate that probably BN and $\theta^1$~Ori~C did not get closer than $\sim$5000~AU in the past. 
%(the minimum separation took place about 4400~years ago).
This large separation is much greater than the gravitational radius that corresponds to the inferred escape speed of BN, which argues against a close encounter. Nevertheless, the accuracy of our measurement (at a 2.3$\sigma$ level) does not conclusively exclude it. 
We note that the uncertainties here are a factor of 5 larger than
quoted by \citet{Gom08}, whose analysis used the relative motion of BN
with respect to Source~I to achieve a greater accuracy. However, since
Source~I has a large absolute proper motion (see Table~\ref{pmt}),
using the BN motion relative to Source~I is undesirable.

Apart from the measurements reported in this paper, there is a number
of pieces of evidence that are inconsistent with an ejection of BN from  $\theta^1$~Ori~C.
First,  $\theta^{1}$~Ori~C and BN have similar absolute radial velocities measured in the ONC rest-frame, $-15$~\kms  \citep{Sta08} and 11~\kms  \citep{Rod09}, respectively, which is inconsistent with  momentum conservation, since BN is approximately five times less massive than $\theta^{1}$~Ori~C.
% momentum conservation would require a much lower blueshifted velocity for $\theta^{1}$ Ori C or much higher redshifted velocity for BN than is observed, if BN has truly been ejected from the $\theta^{1}$ Ori C system in the recent past.
Second, a close passage of BN and $\theta^1$~Ori~C would explain the large proper motion of only BN but not that of Source~I. We note that in order to reach a velocity dispersion of 15~\kms, a stellar cluster on the order of 1000 AU in extent would have to reach 250~\ms, which is greater than the mass concentrations believed to exist in the part of BN/KL local to Source~I.
Third, this scenario would not explain the close passage ($\sim$0\pas1) of BN with Source~I, since it would be a notable coincidence for a high-velocity YSO ejected from a massive binary to pass so close to another high-mass YSO.
Finally, BN exhibits characteristics of a star much younger than any of the Trapezium members, which according to \citet{Hil97} are a few million years old. While none of the Trapezium stars are surrounded by obvious circumstellar material, BN is surrounded by a molecular disk \citep{Sco83, Beu10}  and by a hypercompact HII region with a $\sim$20~AU radius, implying that this star is dragging along a significant amount ($>10^{-4}$~\ms) of dense circumstellar gas \citep{Sco83,Bal05}. 

 %--------------------------------------------------------------------------------------
%--------------------------------------------------------------------------------------

\subsection{Scenario 2: Two-body interaction: star-disk encounter with BN}
\label{double}
We now consider the case where BN and Source~I were initially unbound and approached to within $\sim$50~AU (Sect.~\ref{pas}).  Source~I is surrounded by an ionized disk, indicating a massive (possibly still accreting) protostar (Sect.~\ref{srci}). A close passage by a massive star (BN) should produce tides in the disk and enhance accretion \citep[e.g.][]{Ost94}. 
In particular, \citet{Pfa06} has shown that disks around massive YSOs
located in clusters can experience angular momentum losses of up to
$\sim$50\%-95\% (with consequent increases in accretion), as a result
of gravitational interaction during stellar encounters; this is termed 
{\it cluster-assisted accretion}.

Although star-disk interaction at the center of clusters can affect accretion, it cannot itself lead to stellar ejection. In fact, a two-body interaction does not allow efficient transfer of momentum and/or energy between the two objects. On the contrary, simulations show that disk-assisted capture increases binary formation rates \citep{Moe07}. 
In order to account for the large (positive) total energy of the BN-I system, the formation of a tight binary that develops a large negative-binding energy is required,  a consequence of momentum and energy conservation.
% the total kinetic energy of both I and BN would be  about $10^{47}$ erg, assuming a mass of 10~\ms~and taking 3-D velocities from Table~\ref{pmt}. The accretion of 0.1~\ms~onto a 10~\ms~would then account for this positive kinetic energy, which is not implausible in the presence of a massive (few \ms) envelope. 

In conclusion, in order to end up with stellar ejections, the original system must have contained three or more objects, and one of the successor systems must now be a tight binary.

%-----------------------------------------------------------------------------
 % 
\begin{deluxetable*}{cccccc}
%\begin{deluxetable}{lcccccc}
\tabletypesize{\footnotesize}
\tablewidth{0pc}
\tablecaption{Summary of dynamical scenarios for BN/KL and N-body simulation outcomes. }
\tablehead{
\colhead{Scenario} & \colhead{$M_*$\tablenotemark{a}} & \colhead{$M_{bin}$\tablenotemark{b}} & \colhead{Ejections\tablenotemark{c}} & \colhead{$S_{min}$ Passage\tablenotemark{d}} & \colhead{Sect.} \\ 
\colhead{} & \colhead{(\ms)} & \colhead{(\ms)} & \colhead{(\%)} & \colhead{(AU)} & \colhead{}
}
\startdata
3-body: * * * $\Rightarrow$ * (* *) & 10,10,10 & 10+10 & 0.1  & 7 & \ref{triple} \\
BN, I (new binary) &&&&& \\
&&&&&\\
4-body:  * * * * $\Rightarrow$ * * (* *) & 10,10,5,5 & 5+5 & $<0.1$  & - &  \ref{quadruple} \\
BN, I, n (new binary)\tablenotemark{e} & 10,10,9,1 & 9+1 & $<0.1$  & - & \ref{quadruple} \\
&&&&&\\
11-body: 11$\times$(*) $\Rightarrow$ 9$\times$(*) (* *)  & 0.5($\times$2),1($\times$2),2($\times$2) & 10+10 & 0.2  & 5 &   \ref{cluster} \\
IRcx,BN,n,I (new binary) &3($\times$1),5($\times$1),10($\times$3)&10+5&0.2&4& \ref{cluster}\\
&&&&&\\
3-body: * (* *) $\Rightarrow$ * (* *) & 10,10+10 & 10+10 & 38  & 16 &  \ref{bin} \\
BN(single), I (old binary) &&&&&\\
\enddata
\tablecomments{\\
\tablenotemark{a}{Stellar masses in the N-body simulations.} \\
\tablenotemark{b}{Masses of binary components that remain after ejection of stars from the system.} \\
\tablenotemark{c}{Fraction of cases leading to the ejection of a massive ($\sim$10~\ms) object.} \\
\tablenotemark{d}{Average mimimum passage (and binary separation)  for cases that include an ejection.} \\
\tablenotemark{e}{It assumes that $n$ is a binary with a 5+5 or
  9+1~\ms~total mass, source~I and BN two single 10~\ms~stars.} 
}
\label{scenarios}
\end{deluxetable*}
%\end{deluxetable}
%
%-----------------------------------------------------------------------------

%\subsection{Decay of a multiple ($\ge 3$) system with single objects}
\subsection{Scenario 3: Three and more-body interactions: single stars}
\label{multiple}
In this Section, we describe different scenarios that can result in the high proper motion of multiple stars: a triple system (Sect.~\ref{triple}), a quadruple system (Sect.~\ref{quadruple}), and a stellar cluster  (Sect.~\ref{cluster}). 
In order to test different scenarios and quantify statistical probabilities, we ran N-body simulations using the NBODY0 code from \citet{Aar85}.\footnote{NBODY0 is a direct N-body integrator which for each particle computes the force due to all other N-1 particles. Each particle is followed with its own integration step, an essential feature when the dynamical timescales of different particles in the simulation could vary significantly.}
We numerically integrated several thousand cases of cluster decays with different initial numbers of bodies. Table~\ref{scenarios} summarizes the key aspects and likelihood of different scenarios, which will be described in detail in the following subsections.
We note that each scenario involves the ejection of individual stars and concomitant formation of a tight binary. 
%[Since BN and Source~I are known to be the most massive objects in the region, and the latter is apparently moving with lower velocity than the former, Source~I should be the eccentric massive binary bringing the negative binding energy from the system].
The presence of a disk around Source I poses a challenge for some arrangements (Sect.~\ref{disksurv}). The existence of a binary before the encounter provides an evolutionary path that may agree best with the available data  (see Sect.~\ref{bin}).

\subsubsection{Decay of a triple system  and formation of a new binary}
\label{triple}
In the simplest dynamical scenario of a triple system, we assume that Source~I, which is apparently moving slower than BN, forms an eccentric massive binary, and leaves the system along with BN, as previously proposed by \citet{Rod05} and \citet{Gom08}. %same role as $\theta^1$~Ori~C in the scenario described in Sect.6.1.1.
In order to explain the observed velocities, we need to estimate the mass of both objects. %Mass estimates would go some way to confirming this scenario.
Based on  a luminosity of 2500~\ls~for BN from mid-IR imaging
\citep{Gez98} and  an ionizing photon rate of  $\sim 4 \times
10^{45}~s^{-1}$ from optically thin emission at $\nu = 216$~GHz
\citep{Pla95}, \citet{Rod05} report a mass $M_{\rm BN}=10\pm2$~\ms. In
the following, we will assume a fiducial value of 10~\ms. 
Constraining the mass of Source~I based on an infrared or bolometric
luminosity is hindered by the very high extinction toward the 
source \citep{Gre04b}.
\citet{Mat10} estimate a dynamical mass of $\gtrsim$7~\ms~from SiO masers, whereas \citet{Rei07} report a mass $\approx10M_{\odot}$ based on modeling of proton-electron Bremsstrahlung at 7~mm. 
However, from conservation of linear momentum for the BN-I system, one would estimate a mass:  $M_I=2.3\times M_{BN}\sim 23$~\ms~for $M_{\rm BN}=10$~\ms. This result requires explanation of systematic errors in estimation of the dynamical mass from SiO masers by \citet{Mat10}. We discuss implications in Sect~\ref{mass}.

%CONSERVATION OF ENERGY
Given mass estimates for the stars, the orbital separation of the binary can be estimated from conservation of mechanical energy.
By assuming that the total kinetic energy of the system today, $\frac{1}{2}(M_Iv_I^2+M_{BN}v_{BN}^2)\sim 10^{47}$~erg, is approximately equal to the final binding energy of the binary, $(GM_{Ip}M_{Is})/2a$, where $M_{Ip}$ and $M_{Is}$ are the masses of the primary and secondary stars, respectively, and $a$ is the binary semi-major axis, we obtain $(a/\rm AU) = 40m_f(1 - m_f)(M_{\rm BN}/10~\rm M_{\odot})$, where $m_f$ is the primary's mass fraction in the binary ($m_f=M_{Ip}/(M_{Ip}+M_{Is})$). In the assumption of an equal-mass binary ($m_f = 0.5$), the semi-major axis is 10~AU for $M_{BN}=10$~\ms, corresponding to a period of $\sim$7~yr. %10AU is just the SEMI axis, so that would imply a distance of 20AU among stars. 
 We note that an unequal-mass binary with the same total mass would imply an even smaller orbital separation, 
%but it would have a lower probability of formation and/or survival in a dynamical interaction with other objects 
but it would be dynamically less favorable for ejection  (see discussion in following sections).  

 Hence, a $\sim20$~\ms~equal-mass binary with a semi-major orbital axis $\lesssim 10$~AU will have a binding energy sufficient to compensate the positive kinetic energy of the BN+Source~I system. 
Direct observation of such a tight binary may be beyond what can be achieved with current data. 
Size scales of 10~AU cannot be resolved with the VLA, and no maser
emission has been detected with the VLBA at such small radii
(resulting in the ``dark band" reported by \citealt{Mat10}).
However, highly eccentric orbits may be anticipated to generate local fluctuations in disk heating over just one quarter orbital period ($\sim 2$ years), and these might be recognizable over time. 

 %[In summary, this scenario requires the formation of a tight (O(1 AU) binary with a total mass $\sim 20$~\ms.
The simple analysis above allows estimation of physical parameters of the binary  using basic physics (mass from linear momentum conservation and orbital separation from mechanical energy conservation). In order to quantify the probability of such a triple-body interaction, we simulated a system composed of three single stars of 10~\ms~mass randomly distributed inside a sphere of 1000~AU diameter.  This corresponds approximately to the area subtended by the error cones of proper motions of BN and I (Fig. 5, top panel).
%We found that only 0.8\% of configurations (1000 total) gave ejections with speed higher than 25\%. In all cases with ejections, a 10+10~\ms~binary forms with orbital separations in the range 7-23~AU, with an average value of 15~AU (Table~\ref{scenarios}). The single 10~\ms~star is ejected with a typical velocity $\sim 30$~\kms, while the massive binary leaves the interaction area with a velocity $\sim 15$~\kms, as expected from conservation of linear momentum.
We ran 1000 simulations for durations of 
800 years, adopting a velocity dispersion of 0.4~\kms, corresponding to a thermal velocity in a molecular cloud at $T=10$~K.
We found that only 0.1\% of configurations (1 out of 1000) gave ejections with speed higher than 25~\kms. In the case with ejection, a 10+10~\ms~binary forms with orbital separation $\sim7$~AU (Table~\ref{scenarios}). The single 10~\ms~star is ejected with a velocity $\sim 30$~\kms, while the massive binary leaves the interaction area with a velocity $\sim 15$~\kms, as expected from conservation of linear momentum.
These simple simulations show that the BN-I system is unlikely (probability 0.1\%) to arise from chance encounters among three single stars.

%--------BN AND I +n---------------------------------
%\subsubsection{Decay of a quadruple system: BN,n, and Source I  (new binary)}
\subsubsection{Decay of a quadruple system and formation of a new binary}
\label{quadruple}

Source~I and the BN object may not be the only massive objects in BN/KL.
Radio source ${\it n}$, a YSO associated with a bipolar jet
\citep{Men95} and an accretion disk \citep{Gre04b}, has been proposed
as a possibly significant source of luminosity in BN/KL
\citep{Men95}. Using MIR measurements, \citet{Gre04b} 
estimated a luminosity of $2\times10^3$~\ls, corresponding to a 
stellar mass $\sim8$~\ms. 

Based on position measurements at 8.4~GHz, \citet{Gom08} proposed for source~{\it n} an origin in the cluster decay along with BN and Source~I.
In the following, we investigate this possibility by assuming a proper motion $vx_n=-1.6$~\kms and $vy_n=-21.4$~\kms, as derived by \citet{Gom08} at 8.4~GHz in the ONC rest-frame (but see discussion in Section~\ref{pas} for arguments as to why this measurement is questionable). 
For $M_n=8$~\ms~\citep{Gre04b} and $M_{\rm BN}=10$~\ms~(Sect.~\ref{triple}), conservation of linear momentum (along right ascension and declination) gives a mass  $M_{\rm I}\sim8$~\ms, much lower than that estimated for the triple-system decay (Sect.~\ref{triple}). In this scenario, at least three YSOs (BN, I, and $n$) with comparable masses ($\sim$8-10~\ms) would be running away from each other with high velocities, after a dynamical interaction occurred about 500 years ago.

However, formation of a binary is required to enable ejection of stars
from a cluster.  It is unlikely to be Source~I because an equal-mass
binary of 10~\ms~total would not produce ionizing radiation, while a binary with very different masses for  primary and secondary is an unlikely outcome for an N-body interaction (see below). 
 Hence, in the following analysis, we assume that {\it n} is the binary. A couple of arguments support this hypothesis. First, source~{\it n} has been identified as a hard X-ray source, and since a B-type star is unlikely to be active, $n$ might be an unequal-mass binary with one low-mass member. Second, based on $J$ and $H$ magnitudes,  \citet{Lag04} stated that the SED of source~{\it n} cannot be modeled as a high-mass star with a single extinction value. An additional source of IR luminosity must be added (e.g., circumstellar dust or stellar companion). Although $J$ and $H$ band observations have not revealed bright companion IR sources with separations down to $\simeq$20 AU, a tighter binary remains a possibility. %Hence, if source~{\it n} is a binary, it should have an orbital separation smaller than 20~AU.

In order to estimate the probability that three $\sim10$~\ms~stars are ejected after interaction, we simulated two 4-body interactions that could generate a 10~\ms~binary: 1) two 10~\ms~and two 5~\ms~stars and 2)  two 10~\ms, one 9~\ms,  and one 1~\ms~stars. Initial conditions were set as described in Sect.~\ref{triple}.
Among 1000 simulations, not a single event ends with ejection of a 10~\ms~star and formation of a binary  5+5~\ms~or 9+1~\ms. In system 1) we observed ejections in 0.8\% of cases with masses 5~\ms~(0.7\% of cases) and 10~\ms~(0.1\% of cases). The new binaries formed in those cases had total masses of 10+5~\ms~and 10+10~\ms~in four cases (0.4\% of total) each. Orbital separations varied in the range 2-11~AU (average 9~AU) and 11-23~AU (average 14~AU), for the 10+5~\ms~and 10+10~\ms~binaries, respectively. In system 2) we observed ejections in 2.8\% of cases but only one with mass 9~\ms~(0.1\% of cases), when a binary 10+10~\ms~formed with an orbital separation of 10~AU (the low mass member was ejected in almost all cases).
Even assuming source~{\it n} was a binary before the encounter (with component masses of 5+5~\ms~or 9+1~\ms), an exchange interaction would eject the low-mass companion and form a quasi-equal mass binary of $\sim15-20$~\ms. Hence, unless source~{\it n} is about twice as luminous as believed, the star is unlikely (probability $<0.1$\%) to be the binary that enables ejections from 4-body interaction (Table~\ref{scenarios}).

The conclusion is consistent with other pieces of evidence.
\citet{Gre04b} detected a disk around {\it n} with a radius
$\sim170$~AU, larger than the separation estimated for the BN-I
encounter ($50\pm100$~AU), indicating that either source~{\it n} did
not take part in the encounter at all or that the passage occurred at large separation ($\gtrsim$200~AU). In this last case, source~{\it n} could not have exchanged momentum/energy efficiently, making this hypothesis unlikely. In addition, \citet{Lag04} report detection of a brown dwarf separated by $\sim$300~AU from source~{\it n}. If this is indeed a bound system, then it is unlikely to have survived an earlier dynamical interaction with BN and Source~I.

%The 4-body scenario is motivated by the reported proper motion of source~{\it n} \citep{Gom08}, though the latter should be viewed with a degree of caution. The emission at 8.4~GHz from source~{\it n} is resolved into a bipolar or elongated structure, depending on epoch (size$\sim$0\pas6), and the systematic error in the motion may arise from  internal variability. Tracking of northern and southern components of source~{\it n} by \citet{Gom08} may be possible, but a longer time baseline is required for certainty.  Of some concern, the dataset from \citet{Gom08} is not homogeneous in terms of observing set-up among epochs, and hence changes in uv-coverage could mimic change in structure/position. Using two epochs in which we achieved $5-6\sigma$ detections at 7~mm wavelength, we also measured a proper motion for source~{\it n}, but at $2.7\pm1.4$~mas~yr$^{-1}$ and PA$=67\pm34$ it disagrees with the earlier estimate. In addition, in their Figure~2, \citet{Gom08} show that the displacement of  source~{\it n} in the plane of the sky is approximately half that of BN from the cluster origin.  However, this is inconsistent with the two objects having comparable proper motions and starting their outward motion from the same point of origin and epoch (assuming linearity and constancy of motion). 
%Further observation of source~{\it n} is required to confirm the proper motion measured at 8.4~GHz and hence the dynamical interaction with BN and Source~I. The 7~mm band is preferable because there the emission is compact and modeling errors are thereby reduced.
We conclude that, even assuming that the 8.4~GHz proper motion measurement of source~{\it n} derived by \citet{Gom08} is correct, source~{\it n} is unlikely to have played a major role in the dynamical interaction between Source~I and BN.
%an origin in the cluster decay is questionable.

\subsubsection{Decay of a large-N cluster}
\label{cluster}
To explain the observed motion of BN with respect to Source~I,
\citet{Rod05} proposed an interaction among several members of a
collapsing protostellar cluster. In addition to the radio sources
studied here, many individual peaks of thermal IR ($7-24\mu$m)
emission have been identified in the BN/KL region from high angular
resolution IR studies \citep{Gez98,Gre04b,Shu04}. Some of these may be
embedded, self-luminous protostars  \citep{Gre04b,Shu04},
%the 20 μm brightness contours of the BN-KL area (Gezari et al. 1998) show a morphology strongly reminiscent of that of the Orion Trapezium, particularly as it should have looked at 20 μm before the UV radiation from θ1 Orionis C had ionized and cleared the obscuring envelope. Also, the projected dimensions (d ~ 15'') and the bolometric luminosity of the BN-KL complex (L &sime; 105~\ls) are similar to those of the Orion Trapezium. 
suggesting a protostellar density  even higher than that of the Trapezium cluster. For example, the 5 knots in IRc2 plus Source~I would constitute a cluster core of $10^7-10^8$ pc$^{-3}$ over $\sim 10^3$~AU within the larger ($\sim20''$) IR cluster \citep{Gre04b}. %The large inferred density, greatly exceeding the density of the close-by Trapezium, may be due to cluster youth. As clusters evolve they expand because of stellar dynamics and the expulsion of gas (Kroupa 2004 and references therein). 
Apart from IRc2, also IRc7 and IRc4 (and possibly IRc3) appear to be clustered around the putative expansion center within a few arcseconds (Fig.~\ref{pm}). 

%Hence, the observed motions for the radio sources could be the result of their interaction within a larger protocluster, that also includes some of the IR protostars. 
We propose that the original bound cluster may disintegrate resulting in the expansion of multiple objects with different velocities and the recoil of one (or more) of the protostars as tight binaries. This scenario would explain why the putative center of expansion is presently devoid of infrared sources (see Fig.~\ref{pm}). 
We explicitly note that, owing to the lack of astrometric IR observations, the possibility that the members of the IR cluster in BN/KL are participating in the expansion measured for Source~I and BN, remains speculative. 

%In order to test this scenario, we ran N-body simulations using the NBODY0 code from \citet{Aar85}. We numerically integrated several hundred cases of decaying protostellar clusters with eleven bodies,
In order to test the present scenario, we ran simulations of a protostellar cluster with eleven members, adopting
 masses of 0.5 (2 stars), 1 (2 stars), 2 (2 stars), 3 (1 star), 5 (1 star), and 10 (3 stars)~\ms~(total mass of 45~\ms), randomly distributed inside a sphere of 1000~AU diameter. %This corresponds approximately to the patch on the sky subtended by the error cones of proper motions of BN and I (Fig. 5, top panel). 
The implied stellar density is $1.6\times10^8$~stars~pc$^{-3}$, comparable to that found in the Orion Trapezium and the Cepheus A HW2 stellar complex \citep[e.g.][]{Com07,Jim09}. 
The choice of masses reflects plausible masses for the objects present in the region, based on radio and IR data. The three stars with 10~\ms~are meant represent BN, I, and {\it n}, while the low-to-intermediate mass objects represent the IR  knots associated with IRc2, IRc3, IRc4, and IRc7. 

%We ran simulations for two different velocity dispersions, 0.4 and 1~\kms, corresponding to thermal velocities in a molecular cloud at $T=10$~K and $T=63$~K, for  1000 years each. 
We ran 1000 simulations with durations of 800 years, 
adopting a velocity dispersion of 0.4~\kms. 
%corresponding to a thermal velocity in a molecular cloud at $T=10$~K. 
In a significant fraction of configurations (17\%), stars escaped from
the cluster with velocities greater than that of BN (25~\kms); we
define these as runaway stars. %The mass of escaping stars was in general larger for lower initial dispersion. Assuming a velocity dispersion of 0.4~\kms, 
%We found that 17\% of configurations gave runaway stars with a maximum mass of 10~\ms~in 0.4\%. 
The runaway stars had a maximum mass of 10~\ms~in 0.4\% of cases.
 %By using 1~\kms, we obtained 28\% runaways with 3~\ms~as the highest mass. 
High-mass runaways always required the formation of a binary with 10+5~\ms~(0.2\% of total cases) or 10+10~\ms~total mass (0.2\% of total cases). Orbital separations varied in the range 3-7~AU, with an average value of 5~AU (Table~\ref{scenarios}). 
% The remaining 20\% of runaways not associated with a binary had a mass of 5~\ms. 

In the case that two of the massive objects in the system form a binary and source~{\it n} is a high-mass YSO of 8~\ms~\citep{Gre04b}, the total mass of the cluster can be increased by swapping one of the 0.5~\ms~stars for an 8~\ms~star, keeping the same stellar density (total mass of 52.5~\ms). This  would describe a system where $M_I=10+10$~\ms, $M_{BN}=10$~\ms, and $M_n=8$~\ms.
Simulations of the new system resulted in a similar number of total runaway stars (19\%) of maximum 8-10~\ms~in 0.7\% of cases. Of these, one case out of 7 (0.1\% of total) gave runaways without formation of a binary. 
 New binaries had total masses of 10+5~\ms~in one case (0.1\% of total), 10+8~\ms~in one case (0.1\% of total), and 10+10~\ms~in four cases (0.4\% of total). Orbital separations varied in the range 4-11~AU, with an average value of 6.5~AU. Since results are qualitatively similar to the system with total mass of 45~\ms, we did not report the outcome of the simulations for this case in Table~\ref{scenarios}.

Our simulations show that, without the inclusion of very massive
objects ($\sim$20~\ms) in the cluster, the probability of ejecting
massive members ($\ge 8$~\ms) from the system is very low ($<1$\%). We
note that previous N-body simulations from \citet{Gom08} resulted in a
larger fraction of ejections of objects with 8~\ms~(15\%) because they
assumed several cluster members with masses in the range 8-20~\ms~and
allowed for formation of binaries as massive as 36~\ms. However,
such massive objects cannot readily be identified with any
of the known objects in the BN/KL region.

%\subsection{Caveats with the decay of a multiple system of single stars}
%\label{caveats}

\subsection{Caveat with the decay of a multiple system of single stars: Implications of binary formation on the disk around Source I}
\label{disksurv}
%The typical outcome of a multiple ($\ge3$) system decay (when one or more stellar members are ejected) is the formation of a tight binary among the most massive members of the decaying protocluster. Unless one assumes the presence of one extra (undetected) massive object in the region, Source~I is the natural candidate for the massive binary. 
As argued from the results of N-body simulations for systems of three or more stars, Source I is likely to be a binary comprising stars of comparably high mass.
The formation of a tight binary implies a close passage (O(10 AU) or less), well within the uncertainties in the analysis of proper motion data (see Sect.~\ref{pas}). 
However,
the existence of disks or other structures in close proximity to BN and Source~I poses a challenge to explain. We observe an ionized disk with a $\sim 50$~AU radius around Source~I, while BN has associated dust and (molecular and ionized) gas material in a disk-like structure of tens of AU in radius \citep{Jia05,Rod09}. %possibly up to $\sim$3000~AU \citep{Beu10}; %\citet{Jia05} proposes a disk/outflow axis oriented NE-SW, similar to the elongation of the ionized emission, which would then come from an outflow.
%and source~{\it n} is surrounded by a circumstellar disk $\sim 170$~AU in diameter \citep{Gre04b}.  
Close passages truncate disks larger in radius than the minimum stellar separation or destroy them all together.
In the following, we discuss two possibilities: 1) the original disks are destroyed in the encounter and then rebuilt in the following 500 years; 2) the disks are truncated but survive the encounter. 
%[In principle, disks around BN and Source~n might have been larger in the past and got truncated in the dynamical encounter. The problem becomes more severe for Source~I, which is supposed to have formed a tight binary at the encounter. Massive disk truncation results after the system disintegration, accompanied possibly by large-scale accretion, with a consequent burst of outflow activity. Much of the material collected from the three circumstellar disks may end up in a circumbinary disk around the newly formed binary.]

In the first hypothesis, evidence of a disk around Source~I  today is indicative of a viable rebuilding mechanism fueled by either residual material dispersed from the original disks or interstellar gas in the region. We consider two possible mechanisms for rebuilding the disk: Bondi-Hoyle (B-H) accretion and tidal disruption of a low-mass object. 

%BH accretion
BH accretion is the physical mechanism through which a compact object passing through an ambient medium gains mass \citep{Bon44}.  
%Ambient material is focused into an accretion column toward the star. Most of it accretes onto the star's trailing hemisphere, above which streamlines meet, collisions dissipate energy, and the predominant gas motion is parallel to the star's motion (e.g., Figure~1 of \citealt{Thr-Bal08}).
 Material within a critical radius (BH radius) becomes bound and accreted by the
 protostar. Equating the potential and kinetic energies, we define the
 BH radius $R_{BH}=GM/v^2$, an associated accretion rate
 $\dot{M}_{BH}=\pi\rho R_{BH}^2v$, an accretion timescale
 $t_{BH}=R_{BH}/v$, and a disk mass $M_d=\dot{M}_{BH}/t_{BH}$, where
 $M$ is the compact-object mass and $v$ is the velocity relative to
 the surrounding medium of density $\rho$. Assuming $M=20$~\ms\  (Sect.~\ref{multiple}) 
% COMMENT FROM LDM: SHOULD ONE INSTEAD ADOPT THE MASSES OF THE INDIVIDUAL BINARY COMPONENTS FOR THE BH CALCULATION??
and $v=12$~\kms for Source I (Sect.~\ref{res}), and $n(H_2)=10^7$~cm$^{-3}$ for the molecular gas in the region (as in the hot core), we obtain:  $R_{BH}=120$~AU, $t_{BH}=66$~years, $\dot{M}_{BH}=5\times 10^{-6}$~\msyr, M$_d=0.002$~\ms.
Although the radius estimate is consistent with the observations,
a more sophisticated modeling of BH flow, using 2D and 3D simulations
for both subsonic and turbulent flows \citep[e.g.][]{Kru05,Edg04},
demonstrates that gravitational focusing occurs from the initial scale of $R_{BH}$ to the scale of an irregular accretion pseudo-disk of radius $\lesssim 0.1 \times R_{BH}$ 
%which eventually sets up an accretion disk of size $\lesssim 0.1 \times R_{BH}$ 
 \citep[e.g. see Fig.~1 of][]{Thr-Bal08}. Hence, it appears unlikely that BH accretion would support a disk with comparable radius and symmetry to that observed.  
%We note however that larger $R_{BH}$ may be obtained if the ambient medium is moving in the same direction as Source I. A small difference (less than a factor of 2) in relative velocity would result in a size of the BH flow several times the size of the VLBI structure mapped by \citet{Mat10}.
In addition, \citet{Mat10}  report a lower-limit of $\sim 0.002$~\ms~to the disk mass, as estimated from SiO maser emission, whereas \citet{Beu04} report an upper limit of 0.2~\ms~estimated from the continuum flux density at 345~GHz, indicating that the mass gathered in 500 years via BH would be quite low.
%Apart from the size issue, the mass gathered in 500 years appears to be quite low. \citet{Mat10} estimated a lower-limit to the mass of the disk of $\sim 0.002$~\ms, by assuming a cylindrical geometry ($r= 50$~AU and $h = 14$~AU) and a gas density of $n \sim 10^{10}$~cm$^{-3}$, as required to explain the SiO maser emission. \citet{Beu04} estimated an upper limit of 0.2~\ms~to the gas mass within the circumstellar disk from the flux density at 345~GHz, by assuming a dust temperature of 100 K and a dust opacity index $\beta$ of 2. %Although uncertain, these estimates indicate that BH-accretion may not be a very efficient mechanism for building the disk around Source~I. 
We conclude that BH accretion cannot explain the symmetric morphology, the size, and the mass of the disk we observe around Source I. 
%THE EFFECT OF A PRE-EXIOSTING DISK: \citep{Moe09} performed simulations studying the interaction between a (protoplanetary) disk and a dense ambient medium. The BH accretion onto the star-disk system will not unbind the disk but will redistribute material inwards. This would affect significantly the physical structure of the disk, but not its size (this might actually decrease owing to material redistribution). 
%[A possible way to increase the size of the disk is to apply a shear velocity of the order of the stellar velocity. This would imply a velocity difference of 24~\kms over $\sim 100$~AU, which is unreasonably large.] this corresponds to a velocity gradient of ~10~\kms over a distance of 150 AU: huge. 

%tidal disruption of A LOW-MASS OBJECTS
Another possible mechanism for building the disk around Source~I is suggested by simulations from \citet{Dav06}, which show that a close encounter between 10~\ms~and 3~\ms~stars can lead to a tidal disruption of the low-density 3~\ms~star and subsequent formation of a massive disk around the massive star. If, in the encounter, one low-mass star were stripped away, this process would contribute to create a massive disk/envelope around Source~I, with possible rapid accretion of a significant fraction of it onto the protostar. This scenario, however, faces two major challenges.
%1
First, in order for the tidal disruption to be effective, the two
stars must pass each other to within roughly 
one stellar radius, and by implication, the stellar density of the cluster must be quite high, $2\times10^9$~\ms~pc$^{-3}$ \citep{Dav06}.
%2) 
Second, material following a disruption must be redistributed from a
characteristic radius of a few tenths of an AU (orbital separation of the two stars) to the observed disk radius of tens of AU.
%By equating escape and Keplerian velocity, one can obtain the radius of the disk formed in the disruption event: $r_d=GM/v_k^2 \sim GM/v_s^2$. In order to bring material from 0.1~AU to 50~AU the interloper would need to move in an hyperbolic orbit with $v=500 \times v_k$, which would imply an implausibly large velocity.
The timescale for disk spreading by viscosity is $t_{\nu}\sim
r^2/\nu$, for disk radius $r$ and viscosity $\nu \propto \alpha$
($\alpha$ is the viscosity parameter). We obtain viscous timescales
ranging from $10^5$~yr ($r=10$~AU, $\alpha=10^{-2}$) up to $10^8$~yr ($r=100$~AU, $\alpha=10^{-3}$; \citealt{Hol00}). Tidal disruption is hence an improbable mechanism for the reformation of the disk around Source~I in 500~years.
%ALTERNATIVE, MORE DETAILED CALCULATIONS: $nu=\alpha c_s h$, where $h$ is the disk height, and $\alpha$ is an observational parameter $0.0001\less \alpha \less  1$, $h=c_s/\omega$, $t_d \sim r^2/\nu \sim r^2 / \alpha c_s h \sim r^2 \omega / \alpha c_s^2$. [I NEED A REFERENCE FOR THIS CALCULATION!!!] Assuming $c_s=0.2$~\kms, $r=50$~AU, $\alpha=10^{-2}$, and $\omega=\sqrt{GM/r^3}$,	one would get $10^6$~years for viscous spreading.

We conclude that the observation of a relatively large and symmetric disk around Source~I is inconsistent with the formation of a binary and reformation of a circumbinary disk after 500 years.

%2ND OPTION
We consider then the alternative hypothesis where the original disks survive, although severely affected by the dynamical interaction (e.g., truncation, mass redistribution, accretion burst, etc.).
 In order to assess conditions under which a circumstellar disk is retained in a dynamical encounter or close passage, one should include disk particles in the simulations, which is beyond the scope of this article.  
  A follow-up paper (Moeckel et al. in prep.) will consider a
  dynamical interaction between a binary associated with a disk and a
  single star.  The probability of disk survival in an interaction
  between a binary and a single is higher than a complex three-body
  interaction. The main result is that a significant fraction of the
  disk is retained in the case of either a clean and fast exchange of
  companion in the binary (one or two close passages) or in the case 
of survival of the binary, where the original circumbinary disk is truncated at the radius of the close passage with the single star. %In the following Section, we show that the dynamical interaction of a binary-single has a higher probability. ]

 %--------------------------------------------------------------------------------------
\begin{figure}
\centering
\includegraphics[width=0.5\textwidth,trim=0cm 3cm 5cm 0cm,clip]{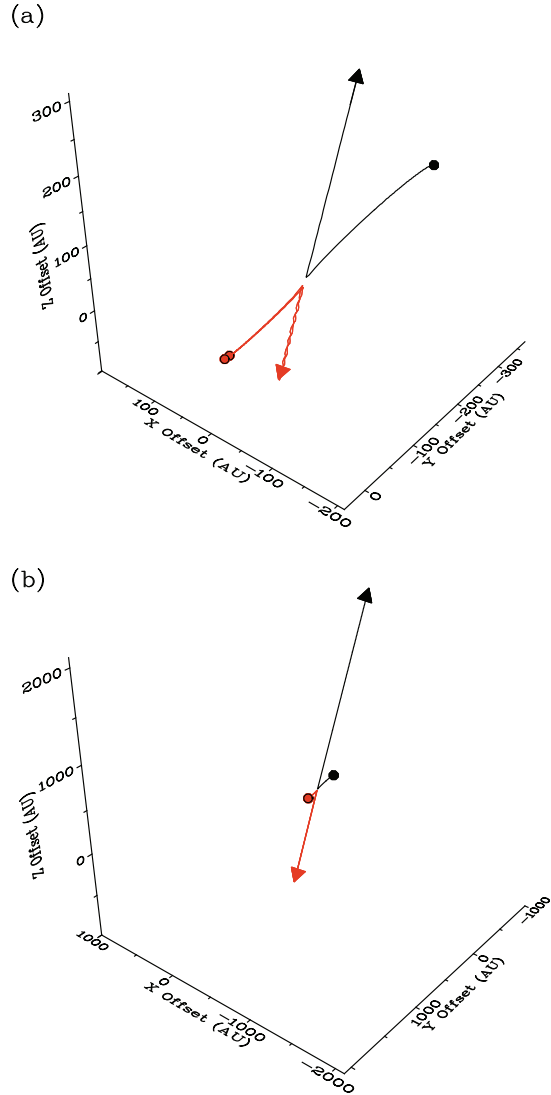}
\caption{
Three body simulation for the interaction between a Source I
binary (10+10 M$_{\odot}$; shown in red) and BN (10 M$_{\odot}$; 
shown in black) at two epochs. Initial positions for the simulations are
marked by filled circles, with the Source I binary components separated by 10 AU and BN 
at a distance of 500 AU. Initial velocities are due to the thermal velocity
of the region, v$_{\rm therm}$=0.4 km~s$^{-1}$.  Arrows mark source positions and 3D motion
directions  at times after closest Source I - BN approach %(separation $\sim$20 AU) 
of {\it(a)} t$\approx$50 years and {\it(b)} t$\approx$500 years. After 500 years, the Source I binary and
BN have positions that are $\sim$3200 AU apart and are moving away from each other with 3D
velocities of 14 and 29~\kms respectively, in agreement with the observations.
}
\label{nbody}
\end{figure}
%--------------------------------------------------------------------------------------

%-------------------------------------------------------------
\subsection{(Favored) Scenario 4: Three-body interaction:  BN and a pre-existing Source~I binary}
%-------------------------------------------------------------
\label{bin}
Ejection of massive stars and formation of tight binaries in multiple-body ($\ge$3) interactions have relatively low probability ($\lesssim1$\%) in the range of stellar densities and masses assumed in our N-body simulations (Sect.~\ref{multiple}). Despite the fact that binaries are  unlikely to arise by chance encounters, observations show that a large fraction ($\gtrsim 60$\%) of massive stars in OB clusters and/or associations are binary and multiple systems  \citep{Mas09}. The abundance of binaries and triples inferred from observations suggests that most stars are born that way  (``primordial" binaries).
Combined hydrodynamic and N-body simulations have shown that primordial binaries can form from both fragmentation of collapsing cores and accretion disks \citep{Ale08,Moe10}. 
 %Near-IR spectral studies have shown that close binaries can form with high fractions ($\gtrsim 50$\% for massive stars)  at a very early stage \citep[e.g.,][]{Pre99,Apa07}.

 %---------------------------------------------------------------
 %THIS IS THE M. KRUMHOLZ SCENARIO
 %Since Source~I has a HII region and a dynamics mass $M_I \gtrsim 7$~\ms~(as estimated from SiO masers), in the following we will assume a binary of total mass of 11~\ms, with mass distributed in 10+1~\ms~or 8+3~\ms~units. %a primary mass $\sim$8-10~\ms~and a secondary mass $\sim$1-3~\ms. 
 %We simulated a 10~\ms~star in parabolic orbit (from infinity) about a binary of total mass 11 solar masses, with mass distributed in 10+1~\ms~or 8+3~\ms~units. The closest approach of the 10~\ms~star to the binary system is set to be 50~AU, as derived from proper motion measurements (Sect.~\ref{pas}).
 %---------------------------------------------------------------
%In the following, we assume that Source~I is a primordial binary which experiences a close passage with the single BN. 
%Section~\ref{disksurv} discusses the shortcomings of scenarios where Source~I forms a binary at the encounter.
In the following, we reconsider a 3-body interaction where two stars lie in a (primordial) binary (Source~I), and a close passage occurs by the third (BN). We assume a binary of total mass 20~\ms, with equal primary and secondary masses of 10~\ms~and an orbital separation of 10~AU. The third star has a mass of 10~\ms~and is at an initial distance of 500~AU. We have numerically integrated 1000 cases with three bodies in a bound system starting with a velocity dispersion $v_{\it therm}=0.4$~\kms~and followed their dynamical interactions for 1000 years. Figure~\ref{nbody} shows the outcome of our simulations for one specific case.
In 39\% of cases, both the single and the binary stars acquire very high speeds in their interaction, flying apart in opposite directions with velocities in the range 10-90~\kms. In all cases, the single 10~\ms~star runs away from the center of interaction with approximately twice the velocity of the binary, in agreement with linear momentum conservation.
In the cases with ejections, the original binary survives in 15\% of cases, while in 24\% of cases a new binary is formed by exchanging one component (39\% total).
The orbital separation of binary components, averaged over the last 50
years of the simulation, varies in the range 1-9~AU, with an average of 4~AU.
The binding energy of the tighter binary (initially at $a=10$~AU)
compensates for the  kinetic energy acquired by the two runaway stars, in agreement with conservation of mechanical energy. 
In the cases where the original binary survives (15\%), the minimum periastron (over 1000 years) between the single and the binary  varied in the range 1-60~AU, with an average value of 16~AU.
% [Figure~x shows the histogram of the impact periastron between intruder and binary, which gives an average value of xAU.] We expect the original circumbinary disk to be truncated to a comparable size, consistent with the observation of a O(50~AU) disk around Source~I. 

%STATE THAT TWO CASES FAILED TO RUN AND WE DIDNT GET RESULTS
The new 3-body simulations show that in a high fraction of cases, massive stars are ejected (39\%) and  the original binary survives (15\%). 
Stellar ejections require a massive binary ($\sim20$~\ms) as initial condition.  In exploring parameter space, we also simulated a 10~\ms~star in parabolic orbit about a binary of total mass 11~\ms, with mass distributed in 10+1~\ms~or 8+3~\ms~units. With these starting conditions, only a small fraction ($\lesssim 1$\%) resulted in ejection, and the star that was lost was the lower mass member.  The process necessarily involved exchange of binary members, where the more massive companion and the third massive object formed an almost equal-mass binary whereas the lower mass member was ejected from the system. 
At the moment, we favor the cases where the binary survives without exchanging components, because therein  the inner reaches of the accretion disk can be preserved (Sect.~\ref{disksurv}). A follow-up paper (Moeckel et al. in prep.) will investigate the effects of a stellar dynamical interaction  on a circumbinary disk as a function of stellar mass, binary orbital separation, single-binary periastron, and swapping of binary members.

\section{Mass of Source~I}
\label{mass}
The assumed dynamical interaction between Source~I and BN, is concomitant with an estimated mass of $\sim 20$~\ms~for Source~I. Recently \citet{Mat10} estimated a dynamical mass using SiO maser proper motion data.  This provided a lower limit of $\gtrsim7$~\ms. %In the following, we investigate implications of an alternate " high-mass" model for source~I.
The influence of non-gravitational forces on gas dynamics may be responsible for the discrepancy between dynamical masses. \citet{Mat10} detected both radial and rotational motions in the inferred disk around the YSO, at radii $< 100$~AU.  Although Doppler velocity data in the outflow show differential rotation of material, the actual rotation curve may be sub-Keplerian, e.g., due to radiative and/or magnetic forces, which can lead to underestimates of the central object mass. In particular, \citet{Mat10} reported possibly bent trajectories among proper motions and suggest that magnetic fields may play a role in shaping the gas dynamics around Source~I.
\citet{Gir09} found recently that the evolution of the gravitational collapse of a massive hot molecular core in G31.41+0.31 is controlled by magnetic fields (on scales $\sim 10000$~AU), which appear to be effective in removing angular momentum and  slowing down circumstellar gas rotation, as expected from magnetic braking \citep{Gal06}. %They show that the rotation velocity decreases with decreasing radius, indicating that the angular momentum is not conserved during the collapse (as expected in a pure Keplerian motion), and proposed magnetic braking as a process to remove the excess of angular momentum \citep{Gal06}. \citet{Mat10} suggested that magnetic fields may play a role in shaping the gas dynamics in Source~I, hinting that magnetic braking might be at work to slow down circumstellar gas rotation, similarly to G31.41+0.31.
\citet{Shu07} considered the case of an accretion disk threaded with magnetic fields that are squeezed in (scales $\lesssim 1000$~AU) from the interstellar medium by gravitational collapse. They find that the poloidal component of the magnetic field produces, as a consequence of its rotation, a change in the radial force balance in the disk, giving a deviation from a Keplerian profile. The resulting rotation law has the same power-law dependence with radius as a Keplerian profile, but scaled with a coefficient $f<1$, which includes the effects of the magnetic field. 
As a consequence of this deviation from Keplerian rotation, the dynamical mass of the central object derived from an observed rotation curve will be smaller than its actual value of a factor $M_{obs}/M_{true}=f^2$, where $M_{true}$ is the true mass and $M_{obs}$ is the observed mass.
 The departure from Keplerian rotation can be analytically expressed in terms of the magnetic coefficient $f$ as:
$1-f^2 = 0.5444M_*/(\lambda_0^2 \dot{M_d} t_{age})$ \citep{Shu07}, where $\dot{M_d}$ is the accretion rate onto the protostar with mass $M_*$, $t_{age}$ is the stellar age, and $\lambda_0$ is a numerical coefficient from the model (assumed to be 4).\footnote{We note that this analysis assumes magnetic fields thread the disk and ignores the stellar magnetosphere. The formula above is then generally applicable to different kinds of stars, i.e. convective (magnetic) and radiative (non-magnetic), or, equivalently, low- and high-mass stars, respectively.}
 Assuming for Source~I $M_{true}\sim20$~\ms (from conservation of linear momentum of the BN-I system) and $M_{obs}\sim8$~\ms (from SiO maser dynamics), we derive $f^2\sim0.4$. Based on the formula above, this value can be readily achieved by assuming reasonable values for mass accretion rates and age: $\dot{M_d}=10^{-5}$~\msyr and $t_{age}=10^5$ years (for $\lambda_0=4$).
This analysis shows that, in the assumption that magnetic fields support the disk, more than half of the mass of the central object can be easily hidden from observations. Nevertheless, we caution that the estimated parameters are indicative and that careful modeling is required to determine the true mass of Source~I.
Future direct measurements of magnetic field amplitude in the disk around Source~I, and an accurate estimate of the mass-accretion rate as well as the stellar age, will enable a better understanding of the effects of magnetic fields on mass measurement from gas dynamics.

\section{What is powering the wide-angle H$_2$ outflow in BN/KL?}
\label{outflow}
The origin, nature, and powering source of the fast, poorly collimated bipolar outflow in BN/KL have been long debated.
The H$_2$ finger structure consists of over 100 individual bow shocks
with a large opening angle and indicates that a powerful event
occurred in the center of BN/KL \citep[e.g.,][]{All93}. Proper motion
studies of Herbig-Haro (HH) objects in BN/KL from the ground
\citep{Lee00} and with the {\it Hubble Space Telescope} \citep{Doi02} showed that more distant bullets are moving faster and indicate an origin about 1000~years ago (assuming no deceleration). \citet{Zap09} observed the CO(2--1) line with the Submillimeter Array and identified several filaments arising from a common center and showing linear velocity gradients with distance from the center.
\citet{Zap09} proposed that the outflow was produced by the disintegration of the young protostellar cluster formed by BN, I, and {\it n}. 

We propose a similar scenario, where the dynamical interaction between BN and the Source~I binary, as described in Sect.~\ref{bin}, may have produced the fast CO outflow and associated H$_2$ fingers in BN/KL.  The outflow has a mass of about 10~\ms~and a kinetic energy of about $3 \times 10^{47}$ ergs \citep{Kwa76}. %If the dynamical process liberated in the surrounding some energy comparable with the kinetic energy of the runaway stars, this could be enough to power the gas explosion visible in the H$_2$ emission lines.
The hardening of a 20~\ms~binary to an orbital separation of 2~AU would release about $5 \times 10^{47}$~ergs of gravitational potential energy, enough to drive the powerful outflow ($3 \times 10^{47}$~ergs) and compensate for the kinetic energy of the BN-I system ($10^{47}$~erg).
This outflow would not be the typical outflow powered by accretion
onto a protostar, but rather the relic of a one-time event.
In this picture, the expanding material was ejected at about the same time and accelerated from the area where the BN-I encounter took place, in agreement with the general ``Hubble-flow'' trend evidenced by proper motions of HH objects and velocity gradients in CO emission  \citep{Zap09}.
%However, there is a discrepancy in the timescales - 1000 years for the finger outflow and dynamical decay (500years), in the assumption of constant velocity in the sky.  ]
Since the  disrupted circumstellar material is dispersed in all directions, this picture explains why the present structure of Source~I is not related to the geometry of the outflow on large scales. 

We note that, although the formation or hardening of a massive binary in a dynamical interaction can explain  energetically the expansion of the outflowing gas, the physical mechanism at its origin remains unknown. More sophisticated combined N-body and hydrodynamic simulations that include also material in circumstellar disks and envelopes may help to understand the fate of circumstellar gas associated with individual protostars participating in a dynamical encounter.

%----------------------------------------------------------------------------

\section{Discussion and Conclusions}
 We have presented high-sensitivity, multi-epoch VLA $\lambda7$~mm continuum observations of the radio sources in the Orion BN/KL star-forming region. These observations have shown that the radio emission from Source~I is elongated NW-SE and is stable over a decade. The morphology and the stability of the emission is consistent with an ionized edge-on disk. 
 
We measured the proper motions of Source~I and the BN object for the first time at 43~GHz, confirming that they are moving with high speeds (12 and 26~\kms, respectively) approximately in opposite direction with respect to each other. We discussed possible dynamical scenarios that can explain the anomalously large motions of both BN and Source~I, and implications for the dynamics of the whole BN/KL region.  Our new  measurements are inconsistent with a close encounter between BN and $\theta^1$~Ori~C in the past, as proposed by \citet{Tan04,Tan08b}.
They confirm the previously proposed scenario of a dynamical interaction between Source~I and BN 500~years ago. In particular, \citet{Gom08} proposed that a dynamical decay of a protostellar cluster resulted in the formation of a tight binary (Source~I) and the ejection of other two sources (BN and {\it n}).
However, the presence of a relatively large, massive disk around Source~I and the low-probability of a multiple-body interaction among single stars, excludes that Source~I formed a binary in the course of the encounter and  subsequently managed to rebuild its disk in 500 years.
We propose that Source~I was a binary system {\it before} the close passage
with BN, which possibly truncated the original circumbinary disk to
$\sim 50$~AU. Inference that this is the natal accretion disk is in agreement with disk-mediated accretion for Source~I, as recently proposed by \citet{Mat10}.
 This dynamical event would have resulted in the ejection of both stars and the hardening of the original binary. 

The dynamical scenarios discussed in this paper were tested with
N-body numerical simulations of dynamical interactions among members
of a protostellar cluster. These simulations, however, illustrate a simplified
physical process of gravitational interaction among point-masses. A
follow-up study will include the effects of a circumbinary disk in the dynamical interactions between protostars.  

%%%%
The mass of Source~I is still controversial. 
From linear momentum conservation of the BN-I system, we 
estimate $\sim20$~\ms---much higher than the value estimated from SiO maser dynamics,  $\gtrsim7$~\ms \citep{Mat10}. On one hand, a strong argument in support of the dynamical estimate is that two independent families of motions,  Keplerian rotation in a disk and escape speed in an outflow,  provide similar mass values, although relevant to different radii and force balance.  On the other hand, the analysis in this paper is strongly based on the fact that 5 to 6 observables (proper motions, LOS velocities, and positions in the plane of the sky) are consistent with the hypothesis of a close passage between Source~I and BN.
One possibility for reconciling the two models is that non-gravitational
forces may play a significant role in driving the dynamics of
molecular gas around Source~I. Assuming the disk is  magnetically
supported, a significant fraction of the mass of the central object
can be hidden in the Keplerian velocity profile of maser emission,
leading to an underestimate of the mass \citep{Shu07}. 
We note that an equal-mass binary with total mass 10~\ms~would not
produce an HII region, and an unequal-mass binary with the same total mass would have a very low-probability of surviving in a dynamical interaction with an equally-massive body (e.g., Sect.~\ref{bin}).
Hence, unless we assume the presence of another very massive source in the region (e.g., {\it n}), the dynamic scenario proposed here suggests a central mass $\sim20$~\ms~for Source~I. We explicitly note that, owing to the non-linear dependence of luminosity on mass, the total luminosity developed by a binary composed of two 10~\ms~stars {\it cannot} explain the high luminosity ($\sim 10^5$~L$_{\odot}$) of the BN/KL nebula. The gravitational energy released by the hardening of a binary with total mass 20~\ms~would, however, be sufficient to power the gas expansion observed in H$_2$ and CO emission in the BN/KL nebula, although a detailed physical mechanism for such a process is still lacking.
%%%%

A final, general remark concerns models of massive star formation. 
Source I is probably the best case known of a massive protostar with ongoing disk-mediated accretion \citep{Mat10}. This work shows that recent dynamical interactions played a fundamental role in shaping the protostellar properties in Orion~BN/KL. Star formation in this region might be then {\it atypical} in the context of the classic theory of isolated low-mass star formation \citep{Shu87}. 
Nevertheless, considering that most (although not all) massive stars are formed in dense protostellar clusters \citep{Lada03}, dynamical interactions in very early stages of cluster evolution may be {\it common} in the context of crowded, massive star formation.
Proper motion studies of the radio/mm continuum sources in similar
regions are required to confirm this hypothesis. The EVLA, when fully
commissioned, will offer broadband (up to 8~GHz) imaging, resulting in a
factor of 10 improvement in continuum sensitivity,  making this kind
of study possible in principle for regions three times more
distant. ALMA, with a similar or even better resolution, will enable
similar studies at (sub)mm wavelengths. Systematic studies with these
instruments will allow us 
to assess the role of dynamical interactions in massive star formation.
%If they turn out to play a dominant role, the uniqueness of the finger signature among star forming regions may suggest it is a transient relic, obscured, or determined by the details of the surrounding ambient material (varying considerably from region to region).

\acknowledgments{We are grateful to Nickolas Moeckel, Mark Krumholz, Leonardo Testi, Daniele Galli, and Malcolm Walmsley for useful discussions.
The data presented here were obtained under the VLA programs AM668A, AG622, AC817, and AC952.
This work was supported by the National Science Foundation under Grant No. NSF AST 0507478.}

\appendix
\label{app}
 %%%%%%%%%%%%%%%%%%%%%%%%%%%%%%%%%%%%%%%%%%%%%%%%%%%%%%%%%%%%%%%%%%%%%  
 \section{Absolute astrometric accuracy}
  The absolute astrometry in each of the program datasets reported
  here has been determined with high accuracy (of order
  $\sim$5~mas). The error at 
each epoch is taken to be the square root of the sum in quadrature of
the 4 main contributions to the total error budget: 
  
 \begin{displaymath}
 e_p = \sqrt{e_t^2+e_n^2+e_s^2+e_{\nu}^2}
 \end{displaymath}
 
\noindent where individual contributions account for angular
separation (calibrator to target), noise limited or fitting
uncertainty, source structure, and frequency-dependent errors. 
Below we describe in detail individual error contributions:
 
 \begin{itemize}
\item
 $e_t$ is the theoretical error in computing the absolute astrometry
 from a calibrator with a separation of $\sim 1^{\circ}$ from the
 target. For a typical baseline length of 10~km and accuracy of 1~cm
 (appropriate for the VLA in A configuration), and a calibrator-target separation of $1.4^{\circ}$, one derives $e_t\sim5$~mas. This is the dominant contribution to the overall error budget. 
   \item
 $e_n$ is the theoretical, noise-limited positional uncertainty,
 given by  $e_n=0.5 \ \theta/SNR$, where $\theta$ is the FWHM of the
 synthesized beamwidth of the array, and SNR is the peak intensity divided by the RMS noise in the map. We compared this error with the fitting error from JMFIT and took the larger of the two (typically  $\sim 1$~mas). 
 
  \item
 $e_s$ is the uncertainty introduced by the source structure. Since both BN and Source~I are resolved at 7~mm in A-configuration, different weighting schemes result in slight changes in  morphology, and possibly in different peak positions. We quantified this contribution by taking the dispersion (i.e., max separation) of the positions in the images with different weighting and restoring beams: $e_s\sim$1~mas for BN and I (resolved sources) and $e_s\sim$0~mas for H and $n$ (point sources). 
  \item
 $e_{\nu}$ is the error introduced by the application of the self-calibration solutions from the maser  to the continuum data, having a frequency separation $\Delta \nu \sim 100-350$~MHz. In previous VLA experiments, we established  an angular offset per MHz as large as 0.01~mas/MHz introduced by calibrating one band by another \citep{God09}. We cannot quantify this contribution for the programs described here (we observed only at one frequency offset from the $v=1$ maser line), but assuming this contribution does not change much from program to program, one can anticipate a contribution at most of a few mas.
 
  \end{itemize}
  
We report typical values of individual contributions in Table~2. By adding up all contributions, typical total uncertainties are $\sim 5$~mas. 
%%%%%%%%%%%%%%%%%%%%%%%%%%%%%%%%%%%%%%%%%%%%%%%%%%%%%%%%%%%%%%%%%%%%%

\end{document}